\documentclass[12pt,english]{article}

\usepackage[T1]{fontenc}
\usepackage[utf8]{inputenc}
\usepackage[dvipsnames]{xcolor}
\usepackage{babel}
\usepackage{amsthm}

\newtheorem{lemma}{Lemma}
\usepackage{amssymb}
\usepackage{mathtools,bbm,dsfont}
\usepackage{mathrsfs}
\usepackage{graphicx}
\usepackage{float}
\PassOptionsToPackage{normalem}{ulem}
\usepackage{etoolbox}
\usepackage{subcaption}
\usepackage{float}
\usepackage{soul}
\usepackage{cancel}
\usepackage{wasysym}
\usepackage{ulem}
\usepackage{braket}

\allowdisplaybreaks[1]

\usepackage{array,longtable}

\makeatletter

\usepackage{jheppub}
\def\@fpheader{\relax}
\allowdisplaybreaks[1]

\makeatother

\begin{document}

\title{Relational quantum dynamics of the black hole interior: singularity resolution and quantum bounce}


\author[a]{Paolo Fragolino,}
\author[a,b,c]{Saeed Rastgoo}
\affiliation[a]{Department of Physics, University of Alberta, Edmonton, Alberta T6G 2G1, Canada} 
\affiliation[b]{Department of Mathematical and Statistical Sciences, University of Alberta, Edmonton, Alberta T6G 2G1, Canada}
\affiliation[c]{Theoretical Physics Institute, University of Alberta, Edmonton, Alberta T6G 2G1, Canada}

\emailAdd{fragolin@ualberta.ca}
\emailAdd{srastgoo@ualberta.ca}









\abstract{We study the interior of the Schwarzschild black hole which is isometric to the Kantowski-Sachs cosmological model, using a fully relational and gauge-invariant quantization framework. The physical Hilbert space is constructed via refined algebraic quantization, and quantum dynamics is recovered through the Page-Wootters formalism with a covariant POVM clock built from one of the two configuration variables, whose Hamiltonian is proportional to the momentum of the said variable. Gauge-invariant relational observables for the area of 2-spheres, the Kretschmann scalar, and the expansion scalar of null geodesic are constructed via group averaging (G-twirl) and evaluated on physical states. We find that the Kretschmann and expansion scalars remain finite throughout the black hole, while the area of 2-spheres is bounded below by a minimum value proportional to the uncertainty in the system variable, which is the other configuration variable distinct from the clock variable. In particular, the expansion scalar vanishes and changes sign at the quantum bounce, establishing a black-hole-to-white-hole transition. These results hold for any general clock whose operator forms a canonical pair with the clock Hamiltonian, and require no specific quantization scheme other than the Schrodinger representation. The singularity resolution emerges directly from relationality, the Heisenberg uncertainty principle, and the structure of the physical Hilbert space.}







\maketitle

\section{Introduction}

The problem of time in quantum gravity is among the most profound
conceptual obstacles in theoretical physics \cite{Kuchar1991,Kuchar1992,PhysRevD.40.2598,Isham:1992ms,Rovelli:1990ph}. In background-independent
theories ---where spacetime geometry is itself a dynamical variable---
there is no preferred external time parameter against which evolution
can be defined \cite{rovelli2004quantum,Smolin:2005cz}. The total Hamiltonian of the system is a sum of first-class
constraints, so physical states are frozen: they are annihilated by
the constraint operators and appear static with respect to any coordinate
time. This is the notorious frozen formalism problem, and it renders
the notion of dynamics in quantum gravity apparently meaningless.

A resolution to this problem is offered by the relational approach
to quantum dynamics, in which time is not an external parameter but
an internal variable: one subsystem of the universe is designated
as a clock, and the evolution of the remaining degrees of freedom
is described relative to it. This idea, first articulated in the context
of quantum cosmology by Page and Wootters \cite{PageWootters1983},
has been substantially developed in recent years into a mathematically
precise framework \cite{H_hn_2021,H_hn_2020,H_hn_2019} which establishes
the equivalence between the Page-Wootters conditional state approach,
the Heisenberg picture of relational observables, and the Dirac quantization
program. Central to this framework is the use of covariant positive
operator-valued measures (POVMs) rather than self-adjoint time operators \cite{H_hn_2021}, which circumvents the obstruction of Pauli's theorem and allows physical
clocks with semi-bounded Hamiltonians.

In parallel, the fate of classical spacetime singularities in quantum
gravity remains a central open question. Classical general relativity
predicts that gravitational collapse leads inevitably to singularities,
where curvature invariants diverge and geodesic completeness fails
---a conclusion codified in the Penrose-Hawking singularity theorems \cite{Penrose:1965,Hawking:1970}. Whether quantum effects resolve these singularities is a question
that various approaches to quantum gravity, including loop quantum
gravity, have addressed with encouraging results \cite{Ashtekar_2006,Bojowald:2001xe, Ashtekar2006}. However, many existing
treatments rely on specific quantization ambiguities or ad hoc modifications
of the dynamics \cite{Ashtekar:2003hd,Corichi:2007tf,Engle:2006ar,Fragomeno_2025}, rather than emerging from a fully relational and
gauge-invariant framework.

This paper addresses both problems simultaneously. We study the interior
of a Schwarzschild black hole \cite{D_Ambrosio_2018,HaggardRovelli2015,Bianchi_2018}, which is isometric to the Kantowski-Sachs
cosmological model and admits a homogeneous minisuperspace description,
using the relational quantization framework of \cite{H_hn_2021} combined with refined algebraic quantization (RAQ) \cite{Giesel_2013}. The RAQ
program provides a mathematically rigorous implementation of Dirac's
quantization procedure for totally constrained systems, constructing
the physical Hilbert space as the space of distributional solutions
to the quantum constraints via a group averaging (rigging map) procedure.

The Kantowski-Sachs Hamiltonian constraint \cite{D_Ambrosio_2018}, after a phase space-dependent
lapse redefinition, takes a separable form $C_{H}=p_{b}+\alpha^{2}/p_{a}$,
which admits a natural deparametrization: the canonical variable $b$
serves as a relational clock with Hamiltonian $H_{C}=p_{b}$, and
the areal radius variable $a$, which controls the area of the 2-spheres
in the interior, plays the role of the system. Since the spectrum
of $\hat{H}_{C}=\hat{p}_{b}$ is all of $\mathbb{R}$, the $b$-clock
is an ideal clock in the sense of \cite{H_hn_2021},
admitting a covariant POVM with Dirac-orthonormal clock states and
a self-adjoint time operator canonically conjugate to $\hat{H}_{C}$.

The central results of this paper are threefold. First, we construct
the physical Hilbert space $\mathscr{H}_{\text{phys}}\cong L^{2}(\mathbb{R},dp_{a})$
via the RAQ rigging map \cite{Giesel_2013}, identify the physical states as distributions
supported on the constraint surface, and show that the clock back-reaction
is encoded in the uncertainty of the physical states. Second, we construct
gauge-invariant relational observables for three physically motivated
quantities ---the area of 2-spheres, the Kretschmann curvature scalar,
and the expansion scalar of null geodesic congruences--- by applying
the G-twirl (group averaging in the adjoint representation \cite{Giacomini_2019,delahamette2021perspectiveneutralapproachquantumframe,Vanrietvelde2020changeof,RevModPhys.79.555}) to kinematical
observables conditioned on the clock reading. Third, we evaluate the
expectation values of these observables on physical states, particularly,
Gaussian states and demonstrate in each case that the classical singularity
is resolved: the area is strictly positive and bounded below by a
Planck-area quantum, and the
expansion scalar remain finite throughout the spacetime, and vanishes and changes sign at the bounce. This signals a transition from a trapped
non-singular black hole interior to an anti-trapped non-singular white
hole geometry.

The paper is organized as follows. In Sec. \ref{sec:Quantum-Clocks}
we review the relational evolution framework, the construction of
quantum clocks from covariant POVMs, and the Page-Wootters conditional
state formalism. Section \ref{sec:BH interior metric} presents the
classical Kantowski-Sachs dynamics and its Hamiltonian reduction.
In Sec. \ref{sec:H-space} we construct the kinematical and physical
Hilbert spaces via RAQ. In Sec. \ref{sec:Quantum-Relational-Observables}
we derive the gauge-invariant relational observables for the area
of 2-spheres, Kretschmann scalar, and expansion scalar, where we establish
the finiteness of the last two and the nonvanishing of the first one
on physical states. In Sec. \ref{sec:gauss-states} we apply these
results to explicit Gaussian states and presents numerical and analytical
evidence for singularity resolution and a black-hole-to-white-hole
transition. Finally, we conclude and summarize our results in Sec.
\ref{sec:Conclusion}.


\section{Relational Evolution, POVMs, and Quantum Clocks}\label{sec:Quantum-Clocks}

\subsection{Relational evolution and physical clocks}\label{subsec:Relational-evolution}

In totally constrained systems, which includes all background-independent
gravitational or gravitation plus matter systems, the total Hamiltonian
of the system is just the sum of first class constraints, which in
turn generate gauge transformations. This makes the entire evolution
just gauge transformations. Thus, (Dirac) observables and physical
states are static under this evolution with respect to any time coordinate
parameter, which is the parameter of the gauge group that generates
such evolution. Hence, the observed dynamics of physical systems must
be recovered relationally: one subsystem must be chosen as a ``clock''
denoted by $C$ to track the evolution of the remaining system denoted
by $S$. For this method to work out, a particularly convenient realization
arises when the Hamiltonian can be cast in a separable or deparametrized
form $H=H_{S}+H_{C}$ (with the same separable form in the quantum
theory). This separability is non-generic and typically requires a
suitable choice of clock variable that leads to a deparametrized description
of the system, where one of the canonical variables (the clock) vanishes
from the Hamiltonian. This is achievable in homogeneous minisuperspace
models. One example of such a model is the interior of a Schwarzschild
black hole, which is isometric to the Kantowski-Sachs cosmological
model, which is homogeneous (but not isotropic). Hence, the method
just described can be applied to study the dynamics of the interior
of a Schwarzschild black hole, and in particular the fate of its singularity,
in the relational approach. We briefly review the classical interior
dynamics in Sec. \ref{sec:BH interior metric} and show that its Hamiltonian
is indeed separable even in the absence of matter fields. In what
follows we keep the discussion general, and in later chapter, we specialize
it to the Schwarzschild interior. 

The idea behind this relational evolution is that, in a frozen timeless
universe, implied by what discussed in the previous paragraph, if
the global state or density operator $\rho$ is heavily entangled,
a subsystem (the clock), described by $\rho_{C}$, can be highly correlated
with another subsystem (the rest of the universe; i.e., the system),
described by $\rho_{S}$. Then, based on Bayes' theorem, if an observer
Alice inside the universe measures the clock and finds it reads $t_{1}$,
because of entanglement, Alice can update her knowledge of the rest
of the universe using quantum conditional probability yielding $\rho_{S}(t_{1})$.
Later, she looks at the clock and sees $t_{2}$, and the conditional
probability yields a state $\rho_{S}\left(t_{2}\right)$ for the system.
By continuously conditioning the static, global state on different
readings of the internal clock, the observer pieces together a sequence
of conditional states. Relational dynamics crucially relies on entanglement
between clock and system; in product states, no nontrivial evolution
can be reconstructed.

In classical statistics, given two random variables, $S$ (System)
and $C$ (Clock), their joint probability distribution is $P(S,C)$.
If one wants to know the state of the system given that the clock
reads a specific time $t$, one uses the formula for conditional probability
\begin{equation}
P\left(S\mid C=t\right)=\frac{P(S,C=t)}{P(C=t)}.\label{eq:baez-general}
\end{equation}
Here, $P(S,C=t)$ is the joint probability of $S$ given that $C$
has value $t$, and $P(C=t)$ is the probability that the clock reads
$t$. Quantum mechanically, the joint probability $P(S,C)$ is the
global entangled density operator $\rho$. This contains all the correlations
(entanglement) between the system and the clock. The joint probability
of $S$ and $C$ given the clock reads $t$, or is in the state $\left|t\right\rangle $,
is 
\begin{equation}
P(S,C=t)=\text{tr}_{C}\left[\left(\hat{\mathbb{I}}_{S}\otimes P_{C}(t)\right)\rho\right].
\end{equation}
The expression $\left(\hat{\mathbb{I}}_{S}\otimes P_{C}(t)\right)\rho$
is the projection of the global state $\rho$ into the clock states
with reading using $t$, where $P_{C}(t)=\left|t\right\rangle \left\langle t\right|$
is the clock projection operator, which is also called a projection-valued
measure (PVM). We will see in the next section, however, that one
needs to use positive operator-valued measures (POVM) instead for
$P_{C}$ due to mathematical reasons. The trace gives us the probability
distribution solely over $S$ given the conditioning on the clock
reading $C=t$. The denominator in \eqref{eq:baez-general} quantum
mechanically becomes
\begin{equation}
P(C=t)=\text{tr}\left[\left(\hat{\mathbb{I}}_{S}\otimes P_{C}(t)\right)\rho\right],
\end{equation}
which is the probability that the clock will be found in state $\left|t\right\rangle $.

\subsection{POVM and quantum clocks}\label{subsec:POVM-and-quantum-clock}

In order to construct the probability formula mentioned in the previous
subsection, we need the clock states, and hence a clock operator $\hat{T}$.
To this end, one might attempt to define a self-adjoint time operator
$\hat{T}=\hat{T}^{\dagger}$ that is canonically conjugate to the
clock's Hamiltonian $\hat{H}_{C}$, such that $\left[\hat{T},\hat{H}_{C}\right]=i\hbar\hat{\mathbb{I}}_{C}$.
However, Pauli's theorem dictates that if a self-adjoint time operator
$\hat{T}$ and a Hamiltonian $\hat{H}_{C}$ satisfy such a global
canonical commutation relation, then both operators must have an absolutely
continuous spectrum equal to $\mathbb{R}$. Because any realistic
physical clock must have a Hamiltonian bounded from below to ensure
a stable ground state, a self-adjoint $\hat{T}$ strictly cannot be
used. Hence, for a realistic physical clock where the Hamiltonian
is bounded from below with a spectrum $\sigma_{c}=\left[\epsilon_{\min},\infty\right)$,
one must restrict the canonical commutation relation to such domain
\cite{nielsen_chuang_2010}. In quantum measurement, a self-adjoint
operator corresponds to a PVM, which relies on a basis of mutually
orthogonal states. Hence, abandoning a self-adjoint $\hat{T}$, means
abandoning PVMs and their orthogonal states. One instead employs POVMs
\cite{Busch1995,H_hn_2021,H_hn_2021_2}. These are bounded positive
semidefinite operators $\hat{E}$ that generalize standard quantum
observables by relaxing the requirement of orthogonality, allowing
for the description of unsharp or realistic measurements.

More concretely, let $\mathscr{H}_{C}$ be the clock Hilbert space,
and let $\hat{H}_{C}$ be the clock Hamiltonian generating a 1-parameter
unitary group of time translations, $\hat{U}_{C}(t)=\exp\left(-\frac{i}{\hbar}t\hat{H}_{C}\right)$.
A quantum clock is formally defined by a POVM map $\hat{E}_{T}:\mathcal{B}(G)\rightarrow\mathcal{B}^{+}\left(\mathscr{H}_{C}\right)$
from the Borel subsets $X$ of the parameter group (typically the
real line, $G=\mathbb{R}$ ) into the space of positive semi-definite
operators on $\mathscr{H}_{C}$. Note that this is not the clock operator.
The clock operator is made of the first moment of the POVM as we will
see below. To function as a valid clock, the POVM must satisfy completeness
to ensure probabilities conserve to unity 
\begin{equation}
\int_{\mathbb{R}}\hat{E}_{T}(dt)=\hat{\mathbb{I}}_{C},\label{eq:Complete-POVM}
\end{equation}
where $dt$ represents the invariant Haar measure of group $G$. The
above also implies that the time states $|t\rangle$ generating the
POVM $\hat{E}_{T}$ span the clock Hilbert space $\mathscr{H}_{C}$.
On the other hand, to ensure that the clock ticks uniformly with respect
to the dynamics generated by $\hat{H}_{C}$, the POVM must satisfy
the covariance condition
\begin{equation}
\hat{U}_{C}(t)\hat{E}_{T}(X)\hat{U}^{\dagger}_{C}(t)=\hat{E}_{T}(X+t),\quad\text{where }t+X\coloneqq\{t+x\mid x\in X\}.\label{eq:Covar-POVM}
\end{equation}
This condition guarantees that translating the clock's state forward
in time is physically equivalent to shifting the measurement apparatus
reading by the exact same amount. The crucial difference between POVMs
and PVMs is that the former are not necessarily orthogonal. Writing
\begin{equation}
\hat{E}_{T}(dt)=\hat{e}(t)dt,\label{eq:E-e}
\end{equation}
this means that 
\begin{equation}
\hat{e}(t)\hat{e}\left(t^{\prime}\right)\neq\delta\left(t-t^{\prime}\right)\hat{e}(t).\label{eq:e-e-not-ortho}
\end{equation}
As a result the time states $|t\rangle$ associated to $\hat{E}_{T}$
form an overcomplete resolution of identity on $\mathscr{H}_{C}$
and are not linearly independent, so they do not comprise a basis
for $\mathscr{H}_{C}$. This can be seen if we write Eq. \eqref{eq:E-e}
more explicitly in terms of clock states $\left|t\right\rangle $
as
\begin{equation}
\hat{E}_{T}(dt)=\hat{e}(t)dt=\mu|t\rangle\langle t|\,dt.\label{eq:Et-ket-bra-mu}
\end{equation}
Replacing the above in \eqref{eq:e-e-not-ortho} yields 
\begin{equation}
\hat{e}(t)\hat{e}\left(t^{\prime}\right)=\mu^{2}|t\rangle\left\langle t\mid t^{\prime}\right\rangle \left\langle t^{\prime}\right|\Rightarrow\left\langle t\mid t^{\prime}\right\rangle \neq\delta\left(t-t^{\prime}\right).\label{eq:e-e-non-ortho-2}
\end{equation}
In fact, this can be shown explicitly if we expand the clock states
in terms of energy eigenstate $\left|\epsilon\right\rangle $. Writing
$\hat{H}_{C}=\int_{\sigma_{C}}d\epsilon\,\epsilon|\epsilon\rangle\langle\epsilon|$
and by restricting the energy spectrum to $\sigma_{c}=\left[\epsilon_{\min},\infty\right)$,
the covariance condition \eqref{eq:Covar-POVM} implies that the time
states can be written in the form
\begin{equation}
\left|t\right\rangle =\int^{\infty}_{\epsilon_{\min}}d\epsilon\,e^{ig(\epsilon)}e^{-\frac{i}{\hbar}\epsilon t}\left|\epsilon\right\rangle ,\label{eq:Clock-state-general}
\end{equation}
where $e^{ig(\epsilon)}$ encodes the phase freedom of the clock states
which parametrizes inequivalent choices of clock. Then computing $\left\langle t^{\prime}\vert t\right\rangle $
and computing the integral by substitution $\omega=\epsilon-\epsilon_{\min}$,
yields
\begin{equation}
\left\langle t^{\prime}\mid t\right\rangle =e^{\frac{i}{\hbar}\epsilon_{\min}\left(t-t^{\prime}\right)}\int^{\infty}_{0}d\omega e^{\frac{i}{\hbar}\omega\left(t-t^{\prime}\right)}=e^{\frac{i}{\hbar}\epsilon_{\text{min}}\left(t-t^{\prime}\right)}\hbar\left(\pi\delta\left(t-t^{\prime}\right)+i\text{PV}\left(\frac{1}{t-t^{\prime}}\right)\right),
\end{equation}
where $\text{PV}\left(\frac{1}{t-t^{\prime}}\right)$ denotes the
Principal Value of the function $\frac{1}{t-t^{\prime}}$. This explicitly
demonstrates \eqref{eq:e-e-non-ortho-2}. In fact for the three cases
of unbounded, semi-bounded, and bounded cases we have \cite{H_hn_2021}
\begin{equation}
\left\langle t|t^{\prime}\right\rangle \coloneqq\chi\left(t-t^{\prime}\right)=\begin{cases}
2\pi\hbar\delta\left(t-t^{\prime}\right), & \sigma_{c}=\mathbb{R}\\
e^{\frac{i}{\hbar}\epsilon_{\text{min}}\left(t-t^{\prime}\right)}\hbar\left(\pi\delta\left(t-t^{\prime}\right)+i\text{PV}\left(\frac{1}{t-t^{\prime}}\right)\right) & \sigma_{c}=(\epsilon_{\text{min}},+\infty)\\
\frac{i\hbar}{t-t^{\prime}}\left(e^{\frac{i}{\hbar}\epsilon_{\text{min}}\left(t-t^{\prime}\right)}-e^{\frac{i}{\hbar}\epsilon_{\text{max}}\left(t-t^{\prime}\right)}\right) & \sigma_{c}=(\epsilon_{\text{min}},\epsilon_{\text{max}})
\end{cases}.\label{eq:t-tprime-3-cases}
\end{equation}
In the special case of a continuous spectrum, \textbf{$\sigma_{c}=\mathbb{R}$}
above, the states form a Dirac-orthonormal family of generalized states.
This would correspond to an ``ideal clock'' \cite{H_hn_2021}. Combining
this property with \eqref{eq:Et-ket-bra-mu} reveals that the POVM
elements $\hat{E}_{T}(dt)$ overlap, which captures the intrinsic
quantum uncertainty and back-reaction of realistic clocks. In other
words, because $\left\langle t\mid t^{\prime}\right\rangle \neq0$
when $t\neq t^{\prime}$, the states ``leak'' into each other and
we cannot sharply define a state perfectly localized in time if we
restrict the available frequencies (energies) to only positive values.
Hence, this non-orthogonality encodes an intrinsic time-energy uncertainty
and limits the sharpness of temporal localization. Another crucial
property of POVMs is their positivity, meaning that their eigenvalues
are positive. In fact, it can be shown that the eigenvalues of discrete
POVMs can be between $0$ and $1$, as opposed to the eigenvalues
of PVMs that are strictly 0 or 1.

Given the $n$th moment of $\hat{E}_{T}(dt)$ defined as
\begin{equation}
\hat{T}^{(n)}=\int_{G}t^{n}\hat{E}_{T}(dt)=\mu\int_{\mathbb{R}}t^{n}\,|t\rangle\langle t|dt,\label{eq:nth-moment}
\end{equation}
we can finally express the physical time (or clock) operator associated
to $\hat{E}_{T}(dt)$ as its first moment 
\begin{equation}
\hat{T}=\hat{T}^{(1)}=\int_{\mathbb{R}}t\hat{E}_{T}(dt)=\mu\int_{\mathbb{R}}t\,|t\rangle\langle t|dt.\label{eq:Clock-Operator}
\end{equation}
The reason behind this is that the POVM $\hat{E}_{T}(dt)$ is an operator-valued
probability measure. Hence, the integral above yields the expectation
value $\left\langle \hat{T}^{(1)}\right\rangle $ which is equal to
the statistical mean of the POVM measurement outcomes. 

Applying the covariance condition \eqref{eq:Covar-POVM} to the clock
operator \eqref{eq:Clock-Operator} we obtain 
\begin{equation}
\hat{U}_{C}(s)\hat{T}\hat{U}^{\dagger}_{C}(s)=\hat{T}-s\hat{\mathbb{I}}_{C}.\label{eq:operator-covariance}
\end{equation}
The canonical commutation relation can be obtained by differentiating
Eq. \eqref{eq:operator-covariance} with respect to $s$ and evaluating
at $s=0$, using $\left.\frac{d}{ds}\left[\hat{U}_{C}(s)\hat{A}\hat{U}^{\dagger}_{C}(s)\right]\right|_{s=0}=-\frac{i}{\hbar}\left[\hat{H}_{C},\hat{A}\right]$,
which yields 
\begin{equation}
\big[\hat{T},\hat{H}_{C}\big]=i\hbar\mathbb{I}_{C}.\label{eq:T-HC-Commut}
\end{equation}
While the above operator $\hat{T}$ is symmetric, it lacks self-adjoint
extensions. Consequently, the canonical commutation relation \eqref{eq:T-HC-Commut}
holds only weakly on a restricted domain of states. In other words,
since we must restrict the domain of $\hat{T}$ on a dense subspace
of states $\mathcal{D}\subset\mathscr{H}_{C}$ whose wavefunctions
vanish at the boundary $\epsilon_{\min}$, the above commutation relation
is only valid for $\psi\in\mathcal{D}$. This resolves the tension
highlighted by Pauli's theorem by abandoning the requirement of a
self-adjoint time operator and instead using a covariant POVM.

For the special case of an ideal clock, the time operator is Hermitian,
$\hat{T}=\hat{T}^{\dagger}$, and it forms a canonically conjugate
pair with the clock's Hamiltonian $\hat{H}_{C}$. Moreover the clock
states are eigenvectors of $\hat{T}$
\begin{equation}
\hat{T}\left|t\right\rangle =t\left|t\right\rangle ,\qquad\text{for }\sigma_{c}=\mathbb{R}.
\end{equation}
The normalization condition in \eqref{eq:t-tprime-3-cases} then yields
\begin{equation}
\mu=\frac{1}{2\pi\hbar}
\end{equation}
for $\sigma_{c}=\mathbb{R}$.

Finally, having established the POVM for our clock, we can explicitly
define the relational evolution of the system in line with what discussed
in subsection \ref{subsec:Relational-evolution}. Given a global physical
state $\rho=\left|\Psi\right\rangle \left\langle \Psi\right|$ with
$\left|\Psi\right\rangle $ belonging to the physical Hilbert space
$\mathscr{H}^{\text{phys}}_{S}\otimes\mathscr{H}^{\text{phys}}_{C}$
and satisfying $\hat{H}\left|\Psi\right\rangle =0$, relational dynamics
is obtained by conditioning this global state on the clock reading
$t$. Using the Page-Wootters formalism \cite{PageWootters1983,H_hn_2021},
this yields the conditional system state as 
\begin{equation}
\rho_{S}(t)=\frac{\operatorname{Tr}_{C}\left[\left(\mathbb{I}_{S}\otimes\hat{E}_{T}(dt)\right)\rho\right]}{\operatorname{Tr}\left[\left(\mathbb{I}_{S}\otimes\hat{E}_{T}(dt)\right)\rho\right]}=\frac{\operatorname{Tr}_{C}\left[\left(\hat{\mathbb{I}}_{S}\otimes|t\rangle\langle t|\right)\rho\right]}{\operatorname{Tr}\left[\left(\hat{\mathbb{I}}_{S}\otimes|t\rangle\langle t|\right)\rho\right]}.
\end{equation}
This defines evolution of the system relative to the clock. By enforcing
the total Hamiltonian constraint $\hat{H}\left|\Psi\right\rangle =0$,
one can show that this conditional state dynamically satisfies the
Schrodinger equation governed by the system Hamiltonian $\hat{H}_{S}$ \cite{PageWootters1983,H_hn_2021},
thereby recovering standard unitary time evolution from a fundamentally
timeless global state.

\section{Black Hole Interior}\label{sec:BH interior metric}

As discussed before, our goal is to apply the machinery of relational
evolution to the interior of the Schwarzschild black hole. Here we
present the necessary ideas, and in particular show that the Hamiltonian
of the model is separable and deparametrized. Hence, the relational
evolution framework can be applied to this system in a rigorous way. 

The general form for the interior of a static spherically symmetric
black hole can be written as \cite{D_Ambrosio_2018} 
\begin{equation}
ds^{2}=-g_{\lambda\lambda}(\lambda)d\lambda^{2}+g_{xx}(\lambda)dx^{2}+g_{\theta\theta}(\lambda)d\Omega^{2}\label{eq:metric-g}
\end{equation}
where $x$ and $\lambda$ are spacelike and timelike coordinate inside
the black hole, respectively. The metric in the standard Schwarzschild
coordinates $t,\,r$ can be derived by the transformation $t=x$ and
$r=\lambda^{2}$. The geometry in the interior of such a black hole
is dynamical and the topology of a $3D$ spatial foliation $\Sigma_{0}$
with fixed time-like coordinate $\lambda=\lambda_{0}$ is $S^{2}\times\mathbb{R}$,
where $\mathbb{R}$ is associated to the spacelike coordinate $x$
and $S^{2}$ to the angular part of the metric. Hence, the interior
can be pictured as a collection of \textit{\emph{cylinders}} having
an increasing proper length and a decreasing surface area as the time-like
coordinate $\lambda$ runs forward \cite{Bianchi_2018}. Notice that such behavior holds
\textit{\emph{before}} the singularity $\lambda=0$ is reached \footnote{Note that the singularity \textit{\emph{inside}} such a black hole
is spacelike, i.e., a spatial hypersurface, not a point or local region
in space.}. We will also consider the spacelike coordinate $x$ to have a finite
range $x\in[x_{\text{min}},x_{\text{max}}]$, i.e. it runs along an
arbitrary finite portion of the cylinder's axis.

Plugging this metric inside the Einstein's field equation yields 
\begin{align}
g_{\lambda\lambda}(\lambda)= & \frac{4\lambda^{4}}{2GM-\lambda^{2}},\\
g_{xx}(\lambda)= & \frac{2GM-\lambda^{2}}{\lambda^{2}},\\
g_{\theta\theta}(\lambda)= & \lambda^{4},
\end{align}
where $M$ is a constant of integration. Because of the singularity
at $\lambda=0$, $\lambda$ ranges from $-\sqrt{2GM}$ to $0$, which
corresponds to the interior of the black hole, i.e., region II of
the Kruskal extension, but cannot be extended beyond $0$. The other
possible range will be $\lambda\in(0,\sqrt{2GM}]$, but because the
metric components in the coordinates used here are singular at $\lambda=0$,
these two possible ranges cannot be glued together. One can now, in
the same coordinate system, perform a change of configuration variables
$(g_{\lambda\lambda},g_{xx},g_{\theta\theta})\rightarrow(N,a,b)$
\cite{D_Ambrosio_2018} 
\begin{equation}
\begin{aligned}g_{\lambda\lambda}= & N\left(\lambda\right)^{2}\frac{a\left(\lambda\right)}{b\left(\lambda\right)}\\
g_{xx}= & \frac{b\left(\lambda\right)}{a\left(\lambda\right)}\\
g_{\theta\theta}= & a\left(\lambda\right)^{2}
\end{aligned}
\label{eq:Nab-variables}
\end{equation}
such that neither of the solutions of $a,\,b,\,N$ are individually
singular at $\lambda=0$ as we will see below. As a result of the
above change of variables 
\begin{equation}
ds^{2}=-N^{2}(\lambda)\frac{a(\lambda)}{b(\lambda)}d\lambda^{2}+\frac{b(\lambda)}{a(\lambda)}dx^{2}+a(\lambda)^{2}d\Omega^{2}.\label{eq:metric-ab}
\end{equation}
where as usual $N$ is the lapse function. Inserting these variables
into the first order action of general relativity yields \cite{D_Ambrosio_2018}
\begin{equation}
S=\frac{v}{4G}\int d\lambda\left(N-\frac{\dot{a}\dot{b}}{N}\right),\label{eq:action}
\end{equation}
where the dot represents derivative with respect to $\lambda$, and
we have defined $v\coloneqq\int^{x_{max}}_{x_{min}}dx$. The solution
to the equations of motion derived from this action are $a=\lambda^{2},\,b=2GM-\lambda^{2},\,N^{2}=4a$,
which as claimed above are non-singular at $\lambda=0$.

Since $N$ is a Lagrange multiplier the only momenta of the model
are $p_{a}=-\frac{v}{4G}\frac{\dot{b}}{N}$ and $p_{b}=-\frac{v}{4G}\frac{\dot{a}}{N}$.
Performing a Legendre transformation leads to the Hamiltonian 
\begin{equation}
H=-\frac{N}{\alpha}\left(\frac{v^{2}}{16G^{2}}+p_{a}p_{b}\right).\label{eq:constr-classical}
\end{equation}
where we have defined $\alpha=\frac{v}{4G}$ for simplicity, with
$[\alpha]=\text{M}=\text{L}^{-1}$ in $c=1$ units.

Clearly, the above Hamiltonian constraint is not separable in $p_{a}$
and $p_{b}$ and we cannot therefore apply the methods described in
Sec. \ref{sec:Quantum-Clocks} to find a quantum clock and define
a relational evolution between $a$ and $b$ degrees of freedom. However,
since $N$ is just a Lagrange multiplier (gauge parameter) and its
choice does not change the classical dynamics, we can redefine it
as 
\begin{equation}
\bar{N}=-\frac{N}{\alpha}p_{a}.
\end{equation}
Classically, this redefinition is globally well-defined and non-singular
because the on-shell momentum $p_{a}=\text{sgn}\left(\lambda\right)\alpha$
(from the EoM) is a constant with a fixed sign along a given physical
trajectory. 
However, $\bar{N}$ now depends on the phase-space variable
$p_{a}$, and at the quantum level such dependencies might introduce operator ordering
ambiguities in $\hat{H}$ and the quantum theories with $N$ and
$\bar{N}$ might not be equivalent. 
The above lapse rescaling leads
to 
\begin{equation}
H=\bar{N}C_{H}=\bar{N}\left(\frac{\alpha^{2}}{p_{a}}+p_{b}\right).\label{eq:WH-Hamilt}
\end{equation}
This form of the Hamiltonian constraint $C_{H}$ is now clearly separable.
The Hamiltonian constraint can now be written as 
\begin{equation}
C_{H}=H_{C}\left(p_{b}\right)+H_{S}\left(p_{a}\right)=p_{b}+\frac{\alpha^{2}}{p_{a}}\label{eq:CH-HC-HS}
\end{equation}
which is in a deparametrized form. Thus the Hamiltonian 
\begin{equation}
H_{C}=p_{b}=-H_{S}=H_{\text{phys}},\label{eq:Class-clock-H}
\end{equation}
generates evolution with respect to internal time $b$. Hence, one
can define a quantum clock out of $b$ using $H_{C}=H_{b}=p_{b}$,
which results in a picture where $a$ evolves with respect to $b$.
We are interested in such an evolution because $a$ is the radius
of the 2-spheres in the interior of the black hole whose areas are
$A=4\pi a^{2}$. Since the classical singularity occurs as $A=0$ \footnote{The Kretschmann scalar $K$ in the Schwarzschild metric is inversely
proportional to the area, i.e. $K\propto\frac{1}{r^{6}}\propto\frac{1}{A^{3}}$.},
we can study the fate of singularity in this way in a relational quantum
theory.


\section{The Hilbert Space }\label{sec:H-space}

In this section we apply the refined algebraic quantization (RAQ)
to our system to find the physical states belonging to the physical
Hilbert space $\mathscr{H}_{\text{phys}}$. RAQ is a mathematically
rigorous framework designed to solve the issue of how to correctly
implement Dirac's quantization program when the constraints have a
continuous spectrum and $0$ is in the spectrum of the operators involved,
which is the case for all totally constrained systems. Since this
procedure is important for the rigor of our construction, we review
it in some detail in Appendix \ref{app:RAQ-Brief}.

\subsection{Kinematical Hilbert Space $\mathscr{H}_{\text{kin}}$}\label{subsec:Kin-space}

\subsubsection*{States:}

According to the RAQ steps discussed in Appendix \ref{app:RAQ-Brief},
the kinematical Hilbert space $\mathscr{H}_{\text{kin}}=\mathscr{H}_{a}\otimes\mathscr{H}_{b}$
of our model consists of square-integrable functions of $p_{a}$ and
$p_{b}$, namely
\begin{equation}
\mathscr{H}_{\text{kin}}=\mathscr{H}_{a}\otimes\mathscr{H}_{b}=L^{2}\left(\mathbb{R}^{2},dp_{a}dp_{b}\right)=\left\{ \psi:\mathbb{R}^{2}\rightarrow\mathbb{C}\vert\:\int_{\mathbb{R}^{2}}dp_{a}dp_{b}\,\psi^{*}\left(p_{a},p_{b}\right)\psi\left(p_{a},p_{b}\right)<\infty\right\} .
\end{equation}

\subsubsection*{Operators:}

We represent the classical algebra and the canonical variables in
the momentum representation as
\begin{align}
\left[\hat{a},\hat{p}_{a}\right]= & i\hbar\hat{\mathbb{I}}, & \left[\hat{b},\hat{p}_{b}\right]= & i\hbar\hat{\mathbb{I}},
\end{align}
and
\begin{align}
\hat{a}\psi= & i\hbar\frac{\partial\psi}{\partial p_{a}}, & \hat{p}_{a}\psi= & p_{a}\psi,\\
\hat{b}\psi= & i\hbar\frac{\partial\psi}{\partial p_{b}}, & \hat{p}_{b}\psi= & p_{b}\psi.
\end{align}
Next, we need to represent the constraint \eqref{eq:WH-Hamilt}. It
is, however, seen that care needs to be taken in representation of
the term $\frac{1}{p_{a}}$ in the constraint, since a naive representation
of its action on kinematical wave functions with non-null support
on $p_{a}=0$ can be ill-defined. Moreover, the operator corresponding
to the unitary representation of the gauge for the rigging map given
by $\hat{U}=e^{-is\hat{C}_{H}}$ will contain infinite powers of the
type $\Big(\widehat{\frac{1}{p_{a}}}\Big)^{n}$. Therefore, to be
a square-integrable space, $\mathscr{H}_{\text{kin}}$ will have to
contain states whose limit for $p_{a}\to0$ decay to $0$ faster than
any power of $\frac{1}{p_{a}}$. The exclusion of states with null
fast decay as $p_{a}\to0$ has also an important physical reason:
it amounts to the exclusion of frozen clocks for which $p_{a}=0$. This is because, from Eq.\eqref{eq:action}, we have $p_a \sim \dot{b} $, i.e., $p_a$ is classically proportional to the rate of change of the clock variable $b$.
The $p_a=0$ exclusion will therefore allow us to have a relational model defined with monotonic
clocks. It will also exclude configurations where the relational Hamiltonian
becomes singular and prevents ill-defined evolution. Later in this
section, we will discuss the frozen-clock claim mathematically. Additionally,
resolving this issue naturally leads to nice states for $\Phi$ as
we will see below. 

The treatment of inverse operators was already discussed in \cite{thiemann2023propertiessmoothdenseinvariant,neuser2023smoothinvariantorthonormalbasis}.
In particular, it was proven in \cite{thiemann2023propertiessmoothdenseinvariant}
that there exist a countable set of wave functions $\{\phi^{(n)}_{\sigma,\tilde{\sigma}}(p_{a})\}_{n\in\mathbb{Z}}$
belonging to the domain of functions that decay at both $p_{a}=\pm\infty$
and $p_{a}=0$ faster than any power of, $p^{n}_{a}$ and $\frac{1}{p^{n}_{a}}$,
respectively. Such states are given by
\begin{equation}
\phi^{(n)}_{\sigma,\tilde{\sigma}}(p_{a})\sim p^{n}_{a}e^{-\frac{p^{2}_{a}}{2\sigma^{2}}}e^{-\frac{\tilde{\sigma}^{2}}{2p^{2}_{a}}},\;\;\;n\in\mathbb{Z},\;\sigma,\tilde{\sigma}\in\mathbb{R}\backslash\{0\},\label{eq:Thiemann-funct}
\end{equation}
defined up to a normalization constant. It can be proven that the
set 
\begin{equation}
D_{0}\left(p_{a}\right)=\text{span}\left\{ \phi^{(n)}_{\sigma,\tilde{\sigma}}(p_{a})|n\in\mathbb{Z},\;\sigma,\tilde{\sigma}\in\mathbb{R}\backslash\{0\}\right\} \label{eq:D0-pa}
\end{equation}
defined in Eq. \eqref{eq:Thiemann-funct} has the following properties
\cite{thiemann2023propertiessmoothdenseinvariant}:

\begin{lemma}\label{lemma}

\leavevmode
\begin{enumerate}
\item $D_{0}$ is an invariant domain for any polynomial in $\hat{p}_{a},\,\widehat{\frac{1}{p_{a}}},\,\hat{a}$. 
\item $D_{0}$ is dense in $L^{2}(\mathbb{R},dp_{a})$. 
\end{enumerate}
\end{lemma}

Property $2$ in Lemma \ref{lemma}, shows that the states \eqref{eq:Thiemann-funct}
generate $L^{2}(\mathbb{R},dp_{a})$. More concretely, the closure
of span of these wave functions is $\mathscr{H}_{a}$ itself, 
\begin{equation}
\mathscr{H}_{a}=\overline{D_{0}\left(p_{a}\right)}=L^{2}(\mathbb{R},dp_{a})\label{eq:Ha-Dpa}
\end{equation}
Using these wave function $\phi^{(n)}_{\sigma,\tilde{\sigma}}$, one
can represent 
\begin{equation}
\widehat{\frac{1}{p_{a}}}\psi=\frac{1}{p_{a}}\psi,\qquad\text{for }\psi\in D_{0}
\end{equation}
due to their property that they fall faster than any power of, $p^{n}_{a}$
and $\frac{1}{p^{n}_{a}}$ at $p_{a}=0$.

As a consequence, the constraint operator in our case is represented
as 
\begin{equation}
\hat{C}_{H}=\hat{p}_{b}+\alpha^{2}\widehat{\frac{1}{p_{a}}},\label{eq:C-H-quantum}
\end{equation}
whose kernel defines the physical states. 

\subsubsection*{Inner product:}

The inner product of this space is the standard one
\begin{equation}
\left\langle \psi_{1}|\psi_{2}\right\rangle _{\text{kin}}=\int_{\mathbb{R}^{2}}dp_{a}dp_{b}\,\psi^{*}_{1}\left(p_{a},p_{b}\right)\psi_{2}\left(p_{a},p_{b}\right).
\end{equation}

\subsection{Seed space $\Phi$ }\label{subsec:phi-space}

\subsubsection*{States:}

Based on the discussion in section \ref{subsec:Kin-space}, and since
$D_{0}(p_{a})$ has the properties mentioned in Lemma \ref{lemma},
the Gelfand triple in our model is
\begin{equation}
\Phi=D_{a}\otimes D_{b}\subset\mathscr{H}_{\text{kin}}\subset D^{*}_{a}\otimes D^{*}_{b}=\Phi^{*},
\end{equation}
where, 
\begin{itemize}
\item $\Phi=D_{a}\otimes D_{b}=D_{0}(p_{a})\otimes S_{0}(p_{b})$, where
$S_{0}(p_{b})$ are Schwartz functions of $p_{b}$. One set of such
Schwartz functions can be given by
\begin{equation}
S_{0}(p_{b})\ni\chi^{(m)}_{k}(p_{b})\sim p^{m}_{b}e^{-\frac{p^{2}_{b}}{2k^{2}}},\;\;\;m\in\mathbb{N},\;k\in\mathbb{R}\neq0\label{eq:S-pb}
\end{equation}
\item $\Phi^{*}=D^{*}_{a}\otimes D^{*}_{b}=D^{*}_{0}\otimes D^{*}_{b}$.
Here $D^{*}_{0}$ is the set of all continuous linear functional on
$D_{a}=D_{0}(p_{a})$ which are tempered distributions acting on the
$\phi^{(n)}_{\sigma,\tilde{\sigma}}(p_{a})$ of Eq. \eqref{eq:Thiemann-funct}.
Moreover, $D^{*}_{b}$ is the space of continuous linear functionals
on $D_{b}=S_0(p_b)$. The space $\Phi^{*}$ is thus the space of all the continuous
linear functionals on $\Phi$ as expected. 
\end{itemize}
As a consequence of the above observations, a generic state $\phi(p_{a},p_{b})\in\Phi$
can be approximated arbitrarily well by finite linear combinations
\begin{equation}
\phi\left(p_{a},p_{b}\right)=N_{\phi}\sum_{m,n}c_{nm}\phi^{(n)}_{\sigma,\tilde{\sigma}}\left(p_{a}\right)\chi^{(m)}_{k}\left(p_{b}\right)\label{eq:state-on-Phi}
\end{equation}
where $N_{\Psi}$ is a suitable normalization function for which 
\begin{equation}
\int_{\mathbb{R}^{2}}dp_{a}dp_{b}\,\phi^{*}\left(p_{a},p_{b}\right)\phi\left(p_{a},p_{b}\right)=1
\end{equation}
and the $c_{nm}\in\mathbb{C}$ are the coefficients of the linear
combination. 

\subsubsection*{Operators:}

As mentioned in step \ref{enu:RAQ-step-2} in Appendix \ref{app:RAQ-Brief},
for an operator $\hat{O}$ in $\mathscr{H}_{\text{kin}}$, the corresponding
operator acting on $\Phi$ denoted by $\hat{O}^{\prime}$ is defined
by restricting its domain to $\Phi$.

\subsubsection*{Inner product:}

To complete our construction of the space $\Phi$, we define its inner
as
\begin{equation}
\left\langle \phi_{1}|\phi_{2}\right\rangle _{\Phi}\coloneqq\int_{\mathbb{R}^{2}}dp_{a}dp_{b}\,\phi^{*}_{1}\left(p_{a},p_{b}\right)\phi_{2}\left(p_{a},p_{b}\right),\quad\forall\phi_{1},\phi_{2}\in\Phi.\label{eq:inner-prod-Phi}
\end{equation}

\subsection{Topological dual $\Phi^{*}$}

\subsubsection*{States:}

As can be seen from step \ref{enu:RAQ-step-3} of RAQ, a typical element
of $\Psi_{\phi}\in\Phi^{*}$ which is a continuous linear functionals,
$\Psi:\Phi\to\mathbb{C}$, on $\Phi$, can be written as
\begin{equation}
\Psi_{\phi}(f):=\langle\phi\mid f\rangle_{\text{kin }}=\int_{\mathbb{R}^{2}}dp_{a}dp_{b}\,\phi^{*}\left(p_{a},p_{b}\right)f\left(p_{a},p_{b}\right).
\end{equation}

\subsubsection*{Operators:}

For a given operator $\hat{O}$ on $\mathscr{H}_{\text{kin}}$ or
equivalently $\hat{O}^{\prime}$ on $\Phi$, the corresponding operator
$\hat{O}$ on $\Phi^{*}$ in our case is
\begin{align}
\left(\hat{O}\Psi_{\phi}\right)[f]\coloneqq & \Psi\left[\hat{O}^{\prime\dagger}f\right],\,\phi,\,\forall f\in\Phi,\Psi\in\Phi^{*}\nonumber \\
= & \int_{\mathbb{R}^{2}}dp_{a}dp_{b}\,\phi^{*}\left(p_{a},p_{b}\right)\hat{O}f\left(p_{a},p_{b}\right),
\end{align}
where we have assumed $\hat{O}^{\prime}$ is self-adjoint and operationally
taken as $\hat{O}=\hat{O}^{\prime}$.

\subsubsection*{Inner product:}

As mentioned in step \ref{enu:RAQ-step-3} of RAQ, no global inner
product can be defined or is needed for $\Phi^{*}$.

\subsection{Physical Hilbert Space}

Owing to the fact that $\mathcal{V}_{\text{phys }}$ inherits the
operators and inner products of $\Phi^{*}$, we start this section
by introducing operators, instead of states.

\subsubsection*{Operators:}

The operators $\hat{O}$ on $\mathcal{V}_{\text{phys }}$, and $\forall f\in\Phi,\,\Psi^{\text{phys}}_{\phi}\in\mathcal{V}_{\text{phys }}$,
can be written as
\begin{align}
\left(\hat{O}\Psi^{\text{phys}}_{\phi}\right)[f]\coloneqq & \Psi^{\text{phys}}_{\phi}\left[\hat{O}^{\prime\dagger}f\right]\nonumber \\
= & \int_{\mathbb{R}^{2}}dp_{a}dp_{b}\,\Psi^{\text{phys}*}_{\phi}\left(p_{a},p_{b}\right)\hat{O}f\left(p_{a},p_{b}\right)
\end{align}
where as in the discussion in the $\Phi^{*}$ space above, we have
assumed $\hat{O}=\hat{O}^{\dagger}=\hat{O}^{\prime\dagger}$. 

\subsubsection*{States:}

As discussed in step \ref{enu:RAQ-step-4} Appendix \ref{app:RAQ-Brief},
the physical states, $\Psi^{\text{phys}}_{\phi}$, constructed out
of seeds $\phi\in\Phi$, are the ones that are in the kernel of the
quantum first class constraints 
\begin{equation}
\left(\hat{C}_{I}\Psi^{\text{phys}}_{\phi}\right)\left[f\right]=\Psi^{\text{phys}}_{\phi}\left[\hat{C}^{\prime}_{I}f\right]=0,\quad\phi,\,\forall f\in\Phi,\,\forall I.
\end{equation}
This leads to the condition
\begin{equation}
\int_{\mathbb{R}^{2}}dp_{a}dp_{b}\,\Psi^{\text{phys}*}_{\phi}\left(p_{a},p_{b}\right)\hat{C}_Hf\left(p_{a},p_{b}\right)=0,
\end{equation}
that should be satisfied by $\Psi^{\text{phys}}_{\phi}$. Here again
we have operationally taken $\hat{C}_H=\hat{C}_H^{\dagger}=\hat{C}_H^{\prime\dagger}$. 

Since in our case, $\hat{U}=e^{-is\hat{C}_{H}},$ these physical states
can be expressed as
\begin{align}
D^{*}_{a}\otimes D^{*}_{b}\ni\Psi^{\text{phys}}_{\phi}\left[f\right]\coloneqq\eta(\phi)[f]= & \frac{1}{2\pi}\int_{G}d\mu(g)\left\langle \hat{U}(g)\phi\mid f\right\rangle _{\mathrm{kin}}\nonumber \\
= & \frac{1}{2\pi}\int_{G}ds\left\langle \hat{U}\left(s\right)\phi\mid f\right\rangle _{\Phi}\nonumber \\
= & \frac{1}{2\pi}\int_{G}ds\left[\int_{\mathbb{R}^{2}}dp_{a}dp_{b}\left(\hat{U}\left(s\right)\phi\left(p_{a},p_{b}\right)\right)^{*}f\left(p_{a},p_{b}\right)\right]\nonumber \\
= & \frac{1}{2\pi}\int_{\mathbb{R}^{2}}dp_{a}dp_{b}\left[\left(\int_{G}ds\hat{U}\left(s\right)\phi\left(p_{a},p_{b}\right)\right)^{*}\right]f\left(p_{a},p_{b}\right)\nonumber \\
= & \int_{\mathbb{R}^{2}}dp_{a}dp_{b}\,\Psi^{\text{phys}*}_{\phi}\left(p_{a},p_{b}\right)f\left(p_{a},p_{b}\right)\label{eq:Psi-Psi-phys}
\end{align}
These are distributional states since $\Psi_{\text{phys}}:\Phi\rightarrow\mathbb{C}$
for all $f\in\Phi$. 

Now, in case of our model, given a seed $\psi\left(p_{a},p_{b}\right)\in\Phi$
and using Eq. \eqref{eq:inner-prod-Phi}, we construct a physical
state in $\mathcal{V}_{\text{phys }}$ as
\begin{equation}
\begin{aligned}\Psi^{\text{phys}}_{\psi}\left[f\right]\coloneqq & \frac{1}{2\pi}\int_{G}d\mu(g)\left\langle \hat{U}(g)\psi\mid f\right\rangle _{\mathrm{kin}}\\
= & \frac{1}{2\pi}\int^{\infty}_{-\infty}ds\left\langle e^{-is\hat{C}_{H}}\psi\bigg|f\right\rangle _{\mathrm{kin}}\\
= & \frac{1}{2\pi}\left\langle \psi\bigg|\int^{\infty}_{-\infty}d\beta\left(e^{-i\beta\hat{C}_{H}}\right)\bigg|f\right\rangle _{\mathrm{kin}},\qquad\beta=-s\\
= & \left\langle \psi\bigg|\delta\left(\hat{C}_{H}\right)\bigg|f\right\rangle _{\mathrm{kin}}\\
= & \int_{\mathbb{R}^{2}}dp_{a}dp_{b}\,\psi^{*}\left(p_{a},p_{b}\right)\delta\left(\hat{C}_{H}\right)f\left(p_{a},p_{b}\right)\\
= & \int_{\mathbb{R}^{2}}dp_{a}dp_{b}\,\left[\delta\left(C_{H}\right)\psi^{*}\left(p_{a},p_{b}\right)\right]f\left(p_{a},p_{b}\right)\\
= & \int_{\mathbb{R}}dp_{a}\,\left[\psi^{*}\left(p_{a},-\frac{\alpha^{2}}{p_{a}}\right)\right]f\left(p_{a},-\frac{\alpha^{2}}{p_{a}}\right),
\end{aligned}
\end{equation}
where we have used the fact that we are in the momentum representation
so $\delta\left(\hat{C}_{H}\right)$ with \eqref{eq:C-H-quantum}
acts by multiplication. Comparing this with \eqref{eq:Psi-Psi-phys}
reveals that $\Psi^{\text{phys}*}_{\psi}\left(p_{a},p_{b}\right)=\psi^{*}\left(p_{a},p_{b}\right)\delta\left(C_{H}\right)$
and hence we find
\begin{equation}
\begin{aligned}\mathcal{V}_{\text{phys }}\ni\Psi^{\text{phys}}_{\psi}\left(p_{a},p_{b}\right)= & \psi\left(p_{a},p_{b}\right)\delta\left(C_{H}\right)\\
= & \psi\left(p_{a},-\frac{\alpha^{2}}{p_{a}}\right)\delta\left(p_{b}+\frac{\alpha^{2}}{p_{a}}\right)\\
= & \psi_{a|b}\left(p_{a}\right)\delta\left(p_{b}+\frac{\alpha^{2}}{p_{a}}\right).
\end{aligned}
\label{eq:phys-states-both}
\end{equation}
Here we have defined the physical amplitude evaluated on the constraint
surface as
\begin{equation}
\psi_{a|b}\left(p_{a}\right)\coloneqq\psi\left(p_{a},-\frac{\alpha^{2}}{p_{a}}\right)\in D_{0}\left(p_{a}\right)\subset L^{2}\left(\mathbb{R},dp_{a}\right),\label{eq:phi-a-b-general}
\end{equation}
which is a linear combination of the states defined in Eq. \eqref{eq:Thiemann-funct}
\begin{equation}
\psi_{a|b}\left(p_{a}\right)=\sum_{n}c_{n}\phi^{(n)}_{\sigma,\sigma^{\prime}}\left(p_{a}\right),\label{eq:physical-state-Thiemann}
\end{equation}
as we will also show in more details in Eq.\eqref{eq:new_thiemanns} below. The ``$a|b$'' index \cite{Vanrietvelde2020changeof,H_hn_2021} indicates
that this physical states describe the evolution of $a$ degrees of
freedom (DoF) with respect to $b$ DoFs. This form completely eliminates
the $p_{b}$-dependence from the evaluated physical amplitude. This
is even more clear if we write $\Psi^{\text{phys}}_{\psi}$ in \eqref{eq:phys-states-both}
in the Dirac bracket notation (with abuse of notation) as 
\begin{equation}
\begin{aligned}\left|\Psi^{\text{phys}}_{\psi}\right\rangle = & \int_{\mathbb{R}^{2}}dp_{a}dp_{b}\,\psi\left(p_{a},-\frac{\alpha^{2}}{p_{a}}\right)\delta\left(p_{b}+\frac{\alpha^{2}}{p_{a}}\right)\left|p_{a}\right\rangle _{a}\left|p_{b}\right\rangle _{b}\\
= & \int_{\mathbb{R}}dp_{a}\,\psi\left(p_{a},-\frac{\alpha^{2}}{p_{a}}\right)\left|p_{a}\right\rangle _{a}\left|-\frac{\alpha^{2}}{p_{a}}\right\rangle _{b}\\
= & \int_{\mathbb{R}}dp_{a}\,\psi_{a|b}\left(p_{a}\right)\left|p_{a}\right\rangle _{a}\left|-\frac{\alpha^{2}}{p_{a}}\right\rangle _{b},
\end{aligned}
\label{eq:physical-states-ket}
\end{equation}
where we have used a notation $\left|\cdot\right\rangle _{b}$, etc.,
to keep track of which space the ket belongs to. We see that in this
expression, the term containing $\left|-\frac{\alpha^{2}}{p_{a}}\right\rangle _{b}$
is redundant. 

Moreover, we can write the state $\Psi^{\text{phys}}_{\psi}\left(p_{a},p_{b}\right)$
in \eqref{eq:phys-states-both} more explicitly in terms of \eqref{eq:Thiemann-funct}
and \eqref{eq:S-pb}, by expressing $\psi$ as in Eq. \eqref{eq:state-on-Phi}
(with $\phi\to\psi$)
\begin{equation}
\Psi^{\text{phys}}_{\psi}\left(p_{a},p_{b}\right)=\psi\left(p_{a},p_{b}\right)\delta\left(C_{H}\right)=N_{\psi}\sum_{m,n}c_{nm}\phi^{(n)}_{\sigma,\tilde{\sigma}}\left(p_{a}\right)\chi^{(m)}_{k}\left(-\frac{\alpha^{2}}{p_{a}}\right)\delta\left(p_{b}+\frac{\alpha^{2}}{p_{a}}\right)\label{eq:phys-state-distrib}
\end{equation}
In the same way and using Eqs. \eqref{eq:state-on-Phi} (with $\phi\to\psi$),
\eqref{eq:D0-pa}, and \eqref{eq:S-pb}, the amplitude $\psi_{a|b}\left(p_{a}\right)$
in \eqref{eq:phi-a-b-general} can be written as
\begin{equation}
\begin{aligned}\psi_{a|b}\left(p_{a}\right)= & N_{\psi}\sum_{m,n}c_{nm}\phi^{(n)}_{\sigma,\tilde{\sigma}}\left(p_{a}\right)\chi^{(m)}_{k}\left(-\frac{\alpha^{2}}{p_{a}}\right)\\
\propto & \sum_{m,n}c_{nm}p^{n}_{a}e^{-\frac{p^{2}_{a}}{2\sigma^{2}}}e^{-\frac{\tilde{\sigma}^{2}}{2p^{2}_{a}}}\left(-\frac{\alpha^{2}}{p_{a}}\right)^{m}e^{-\frac{\alpha^{4}p^{-2}_{a}}{2k^{2}}}\\
= & \sum_{m,n}c_{nm}\left(-\alpha^{2}\right)^{m}p^{n-m}_{a}e^{-\frac{p^{2}_{a}}{2\sigma^{2}}}e^{-\frac{p^{-2}_{a}}{2}\left(\tilde{\sigma}^{2}+\frac{\alpha^{4}}{k^{2}}\right)}\\
:= & \sum_{m,n}c_{nm}\left(-1\right)^{m}\left(\alpha\right)^{2m}p^{n-m}_{a}e^{-\frac{p^{2}_{a}}{2\sigma^{2}}}e^{-\frac{\sigma^{\prime2}}{2p^{2}_{a}}}\\
\propto & \sum_{m,n}c^{\prime}_{nm}\phi^{(n-m)}_{\sigma,\sigma^{\prime}}\left(p_{a}\right),
\end{aligned} \label{eq:new_thiemanns}
\end{equation}
where we have defined
\begin{equation}
\sigma^{\prime2}\coloneqq\tilde{\sigma}^{2}+\frac{\alpha^{4}}{k^{2}}.
\end{equation}
We will comment on this important observation in Sec. \ref{subsec:Clock-Hilbert-space}. 

A similar procedure applies if we decide to solve our delta $\delta\left(p_{b}+\frac{\alpha^{2}}{p_{a}}\right)=\frac{\alpha^{2}}{p^{2}_{b}}\delta\left(p_{a}+\frac{\alpha^{2}}{p_{b}}\right)$
for $p_{a}$ instead of $p_{b}$. In this case, the new wave-function
will be $\psi\left(-\frac{\alpha^{2}}{p_{b}},p_{b}\right)\coloneqq\psi_{b|a}\left(p_{b}\right)$
and the $p_{a}$ dependence will be completely removed. Then the $a$
DoFs would be the clock, with respect to which the $b$ DoFs evolve.
This leads to an important observation: the Hilbert space $\mathscr{H}_{\text{phys}}$
contains all the possible viewpoints at once. In this sense, this
model is fully relational and $\mathscr{H}_{\text{phys}}$ is often
denoted as a Perspective neutral space \cite{Vanrietvelde2020changeof,delahamette2021perspectiveneutralapproachquantumframe,H_hn_2021}.

\subsubsection*{Inner product:}

From step \eqref{enu:RAQ-step-4} in Appendix \eqref{app:RAQ-Brief}
we can see that, using Eq. \eqref{eq:inner-prod-Phi}, in our case
the inner product on $\mathcal{V}_{\text{phys }}$ becomes
\begin{equation}
\begin{aligned}\left\langle \Psi^{\text{phys}}_{\psi}\vert\Psi^{\text{phys}}_{\phi}\right\rangle _{\text{phys}}= & \left\langle \psi\bigg|\delta\left(\hat{C}_{H}\right)\bigg|\phi\right\rangle _{\text{kin}}\\
= & \int_{\mathbb{R}^{2}}dp_{a}dp_{b}\,\psi^{*}\left(p_{a},p_{b}\right)\delta\left(\hat{C}_{H}\right)\phi\left(p_{a},p_{b}\right)\\
= & \int_{\mathbb{R}^{2}}dp_{a}dp_{b}\,\psi^{*}\left(p_{a},p_{b}\right)\delta\left(C_{H}\right)\phi\left(p_{a},p_{b}\right)\\
= & \int_{\mathbb{R}}dp_{a}\,\psi^{*}_{a|b}\left(p_{a}\right)\phi_{a|b}\left(p_{a}\right)
\end{aligned}
\label{eq:phys-inner}
\end{equation}
Finally, by eliminating $p_{b}$ from the expressions and using the
physical states \eqref{eq:phys-state-distrib} and their inner product
\eqref{eq:phys-inner}, we can express the physical Hilbert space
as
\begin{equation}
D_{0}\left(p_{a}\right)\subset\mathscr{H}_{\text{phys}}=L^{2}\left(\mathbb{R},dp_{a}\right)\subset D^{*}_{0}\left(p_{a}\right).\label{eq:D-pa}
\end{equation}

\subsection{Clock Hilbert space $\mathscr{H}_{C}$ and black hole quantum clock}\label{subsec:Clock-Hilbert-space}

We will now construct the quantum clock in our model based on the
instruction in Sec. \ref{subsec:POVM-and-quantum-clock}. In our system,
according to \eqref{eq:Class-clock-H}, the clock Hilbert space is
associated to $b$ DoF, and hence we have
\begin{equation}
\mathscr{H}_{C}\coloneqq\mathscr{H}_{b}=L^{2}\left(\mathbb{R},dp_{b}\right)
\end{equation}
Furthermore, due to \eqref{eq:Class-clock-H}, the clock's Hamiltonian
is 
\begin{equation}
\hat{H}_{C}=\hat{p}_{b},\label{eq:HC-pb-quantum}
\end{equation}
and its eigenstates are the momentum eigenstates \cite{H_hn_2021}
\begin{equation}
\hat{H}_{C}\left|p_{b}\right\rangle _{C}=p_{b}\left|p_{b}\right\rangle _{C}.\label{eq:quant-clock-H}
\end{equation}
Remember that, in the representation used so far, the momentum states
$\left\{ \left|p_{a}\right\rangle _{a}\right\} $ and $\left\{ \left|p_{b}\right\rangle _{b}\right\} $
\textit{\emph{form a Dirac-orthonormal family of generalized eigenstates
basis}} for $\mathscr{H}_{a}$ and $\mathscr{H}_{b}$, respectively,
i.e., $\left\langle p^{\prime}_{a}\vert p_{a}\right\rangle _{a}=\delta\left(p^{\prime}_{a}-p_{a}\right)$
and $\left\langle p^{\prime}_{b}\vert p_{b}\right\rangle _{b}=\delta\left(p^{\prime}_{b}-p_{b}\right)$.

From Eq. \eqref{eq:quant-clock-H}, it is seen that the clock's spectrum
is therefore given by all the possible values of $p_{b}$, i.e., $\epsilon=p_{b}$
and $\sigma_{c}=\mathbb{R}$ (see Eq. \eqref{eq:t-tprime-3-cases}).
This means that the $b$ DOF of the interior of a black hole serves
as an ideal clock \cite{H_hn_2021}, as discussed in Sec. \ref{sec:Quantum-Clocks}.
Hence, according to Eq. \eqref{eq:Clock-state-general} (but with
$\epsilon_{\min}\to-\infty$) the states for this clock are given
by
\begin{equation}
|t\rangle=\int^{\infty}_{-\infty}dp_{b}\,e^{ig\left(p_{b}\right)}e^{-\frac{i}{\hbar}p_{b}t}\left|p_{b}\right\rangle .\label{eq:t-ket-ideal}
\end{equation}
In the specific case of $g(p_{b})=0$, comparing the above with the
Fourier transform $|b\rangle=\frac{1}{\sqrt{2\pi\hbar}}\int dp_{b}\,e^{-\frac{i}{\hbar}p_{b}b}\left|p_{b}\right\rangle $
leads to the conclusion that the clock's states are directly proportional
to $b$, namely 
\begin{equation}
\left|t\right\rangle =\sqrt{2\pi\hbar}\left|b\right\rangle .\label{eq:t-ket-b-ket}
\end{equation}
In this case, from Eq. \eqref{eq:Clock-Operator} and the spectral
theorem, we obtain
\begin{equation}
\hat{T}|_{g=0}=\frac{1}{2\pi\hbar}\int dt\,t|t\rangle\langle t|=\int db\,b|b\rangle\langle b|=\hat{b},\label{eq:T-g0-b}
\end{equation}
and therefore 
\begin{equation}
\hat{T}|_{g=0}\left|t\right\rangle =\hat{b}\left(\sqrt{2\pi\hbar}\left|b\right\rangle \right)=\sqrt{2\pi\hbar}b\left|b\right\rangle .
\end{equation}
As mentioned before in Sec. \ref{subsec:POVM-and-quantum-clock},
in general, the function $g(p_{b})$ encodes the freedom in choices
of inequivalent clocks or equivalently operators $\hat{T}$, which
is any operator that satisfies the commutation relation \eqref{eq:T-HC-Commut}
with $\hat{H}_{C}=\hat{p}_{b}$. In particular, we have
\begin{equation}
\begin{aligned}\hat{T}= & \frac{1}{2\pi\hbar}\int_{\mathbb{R}}dt\,t\left|t\right\rangle \left\langle t\right|\\
= & \frac{1}{2\pi\hbar}\int dt\int dp_{b}\int dp^{\prime}_{b}\,te^{\frac{i}{\hbar}\left(p^{\prime}_{b}-p_{b}\right)t}e^{ig\left(p_{b}\right)}\left|p_{b}\right\rangle \left\langle p^{\prime}_{b}\right|e^{-ig\left(p^{\prime}_{b}\right)}\\
= & \frac{1}{2\pi\hbar}\int dt\int dp_{b}\int dp^{\prime}_{b}\,te^{\frac{i}{\hbar}\left(p^{\prime}_{b}-p_{b}\right)t}e^{ig\left(\hat{H}_{C}\right)}\left|p_{b}\right\rangle \left\langle p^{\prime}_{b}\right|e^{-ig\left(\hat{H}_{C}\right)}\\
= & e^{ig\left(\hat{H}_{C}\right)}\left(\frac{1}{2\pi\hbar}\int dt\,t\int dp_{b}\int dp^{\prime}_{b}\,e^{\frac{i}{\hbar}\left(p^{\prime}_{b}-p_{b}\right)t}\left|p_{b}\right\rangle \left\langle p^{\prime}_{b}\right|\right)e^{-ig\left(\hat{H}_{C}\right)}\\
= & e^{ig\left(\hat{H}_{C}\right)}\hat{T}|_{g=0}e^{-ig\left(\hat{H}_{C}\right)}
\end{aligned}
\label{eq:freedom-of-g}
\end{equation}
Applying the Hadamard Lemma to above yields 
\begin{equation}
\hat{T}=\hat{T}|_{g=0}+i\left[g\left(\hat{H}_{C}\right),\hat{T}|_{g=0}\right]+\frac{i^{2}}{2!}\left[g\left(\hat{H}_{C}\right),\left[g\left(\hat{H}_{C}\right),\hat{T}|_{g=0}\right]\right]+\ldots\label{eq:T-BCH}
\end{equation}
Taylor expanding $g(\hat{H}_{C})$ and using \eqref{eq:HC-pb-quantum}
and \eqref{eq:T-g0-b} leads to \cite{H_hn_2021}
\begin{equation}
\begin{aligned}\left[g\left(\hat{H}_{C}\right),\hat{T}|_{g=0}\right]= & \left[\sum^{\infty}_{n=0}\frac{g^{(n)}(0)}{n!}\hat{H}^{n}_{C},\hat{T}|_{g=0}\right]\\
= & \sum^{\infty}_{n=1}\frac{g^{(n)}(0)}{n!}(-i\hbar)n\hat{H}^{n-1}_{C}\\
= & -i\hbar\sum^{\infty}_{n=0}\frac{g^{(n+1)}(0)}{n!}\hat{H}^{n}_{C}\\
= & -i\hbar\sum^{+\infty}_{n=0}\frac{h^{(n)}(0)}{n!}\hat{H}^{n}_{C}\\
= & -i\hbar\,h\left(\hat{H}_{C}\right),
\end{aligned}
\end{equation}
where we have defined $h(p_{b})\coloneqq\frac{dg(p_{b})}{dp_{b}}$.
This means that all the higher order commutators in Eq. \eqref{eq:T-BCH}
are identically $0$ since any two functions of the same operator
commute. This allows us to rewrite Eq. \eqref{eq:T-BCH} as
\begin{equation}
\hat{T}=\hat{T}|_{g=0}+\hbar\,h\left(\hat{H}_{C}\right)=\hat{b}+\hbar\,h\left(\hat{H}_{C}\right),\qquad h(p_{b})\coloneqq\frac{dg(p_{b})}{dp_{b}}.\label{eq:T-Tg0}
\end{equation}
This implies that the function $g(p_{b})$ inside Eq. \eqref{eq:t-ket-ideal}
has the role of shifting the clock operator $\hat{b}$ by a generic
function $h\left(\hat{H}_{C}\right)$. As a consequence, there exists
an infinite number of clocks $\hat{T}$ that satisfy the commutation
relation \eqref{eq:T-HC-Commut} with the clock Hamiltonian $\hat{H}_{C}=\hat{p}_{b}$,
in which the ``seed'' clock is $\hat{b}$. These clocks correspond
to different choices of time zero and different weightings of energy
eigenstates in the clock state superposition. They are physically
inequivalent in the sense that their POVMs assign different probabilities
to the same measurement outcomes. 

Having constructed the quantum clock, we can make some important observations
about it from the properties of the Hilbert space. One can see that
the only difference between the dense countable set of states \eqref{eq:Thiemann-funct}
and the ones in \eqref{eq:physical-state-Thiemann} that are used
as seeds to construct the physical states, is in their uncertainties,
which changes from $\tilde{\sigma}$ to $\sigma^{\prime}$. This can
be interpreted as the clock's back-reaction, in which the clock's
uncertainty (in our case $\Delta b$) influences the state of the
system (in our case, any observable of $(a,p_{a})$). This can be
seen as follows: in the Gaussians of Eq. \eqref{eq:S-pb}, the uncertainty
in $b$ is proportional to $\Delta b\sim \hbar/k$. Therefore, the new
uncertainty $\sigma^{\prime}$ in $p_{a}$ can be rewritten as
\begin{equation}
\sigma^{\prime2}\sim\tilde{\sigma}^{2}+\frac{\alpha^4}{\hbar^2}(\Delta b)^{2}.\label{eq:clock-back-reaction}
\end{equation}
This clearly shows how the clock's back-reaction influences our measurements
on the system. In particular, $\tilde{\sigma}$ and $\sigma^{\prime}$
regulate the ``suppression'' of the Thiemann's states \eqref{eq:Thiemann-funct}
around $p_{a}=0$, i.e., they encode how fast the probability goes
to $0$ as $p_{a}\to0$. The relation \eqref{eq:clock-back-reaction}
clearly shows that such suppression always gets bigger as the clock's
uncertainty increases. In particular, the clock's uncertainty can
never reduce $\sigma^{\prime}$ to $0$, i.e., it can never contribute
to obtain states that are not rapidly decaying at $p_{a}=0$ for any
powers of $p_{a}$, which is clear from the positivity of the clock's
uncertainty term $\frac{\alpha^{4}}{k^{2}}$ in \eqref{eq:clock-back-reaction}.

It is also important to notice that because of the classical relation
$p_{a}\sim\dot{b}$, $p_{a}=0$ will correspond to a frozen clock,
which is an undesired property for a relational model. In particular,
Eq. \eqref{eq:clock-back-reaction} shows that, the bigger the clock's
uncertainty $\Delta b$, the greater the suppression in $p_{a}=0$,
which suppresses contributions from configurations where the clock
would be effectively frozen. Therefore, the exclusion of states that
are not rapidly-enough decaying at $p_{a}=0$ is justified physically
and is necessary to have a well-defined relational theory that employs
monotonic clocks.

\section{Quantum Relational Observables\label{sec:Quantum-Relational-Observables}}

As we mentioned before, our goal in this paper is to study the fate
of singularity of a Schwarzschild black hole in the relational approach.
To this end, we study three relevant quantities. The first of these
quantities is the area of 2-spheres inside the black hole. Classically, using the metric \eqref{eq:metric-ab}, this area is given by 
\begin{equation}
A=\int d\Omega\sqrt{g_{\theta\theta}g_{\phi\phi}}=4\pi a^{2}.\label{eq:Area-a-class}
\end{equation}
Although the nonvanishing of this area (and in fact its gauge invariant
version as we will see below) may not necessarily mean that the singularity
is resolved, it is a good indicator of such a phenomenon. 

The second quantity which is intimately related to the singularity
resolution is the invariant Kretschmann scalar
\begin{equation}
K=R_{\mu\nu\rho\sigma}R^{\mu\nu\rho\sigma}.
\end{equation}
This is perhaps the most common Riemann invariant used to determine
whether a spacetime is singular and its finiteness across the whole
spacetime signals the lack of any singularity in that spacetime \cite{Bosso:2023fnb}.

The third quantity we consider is the expansion scalar $\vartheta=\nabla_{\mu}k^{\mu}$
of null geodesics described by an affinely parametrized null vector
field $k^{\mu}$. This quantity describes how much the geodesics congruence
focus or defocus in their motion through spacetime. The expansion
scalar is a powerful indicator in detection of singularities and is
the backbone of Penrose-Hawking singularity theorems. If this quantity
never diverges, particularly to $-\infty$, it is a definitive sign
that the spacetime is singularity-free \cite{Rastgoo:2022mks, Hergott:2025elg, Hergott:2022hjm}.

\subsection{General treatment of gauge-invariant observables}\label{subsec:General-treatment-obs}

In this subsection, we will summarize some of the results about gauge-invariant
observable that we use in the rest of this paper. Consider a gauge-dependent
operator (kinematical observable) $\hat{f}_{S}$ associated to the
system $S$ with a Hamiltonian operator $\hat{H}_{S}$, clock Hamiltonian
$\hat{H}_{C}$ with a clock operator $\hat{T}$ and clock eigenstates
$\left|t\right\rangle $ with eigenvalues $t$. We are considering
a general clock $\hat{T}=\hat{T}|_{g=0}+\hbar\,h\left(\hat{H}_{C}\right)$
where $h(\epsilon)=\frac{d}{d\epsilon}g(\epsilon)$ whose states are \cite{H_hn_2021}
\begin{equation}
\left|t\right\rangle _{g}=e^{ig\left(\hat{H}_{C}\right)}|t\rangle_{g=0}=\int^{\infty}_{-\infty}dp_{b}\,e^{ig\left(p_{b}\right)}e^{-\frac{i}{\hbar}p_{b}t}\left|p_{b}\right\rangle _{b},\label{eq:generic-clock-states-g-nonzero}
\end{equation}
where the coefficients of the Fourier transform, or the plane waves
in this case, are 
\begin{equation}
_{b}\left\langle p_{b}|t\right\rangle =e^{ig\left(p_{b}\right)}e^{-\frac{i}{\hbar}p_{b}t}.\label{eq:Fourier-b-t-g-}
\end{equation}
The gauge-dependence of $\hat{f}_{S}$ dictates
\begin{equation}
\left[\hat{f}_{S},\hat{C}_{H}\right]\neq0.
\end{equation}
To find out what the gauge-dependent system observable $\hat{f}_{S}$
looks like when the clock $T$ reaches the target time $\tau$, we
must evolve the system from $t$ to $\tau$ as
\begin{equation}
e^{\frac{i}{\hbar}\hat{H}_{S}(\tau-t)}\hat{f}e^{-\frac{i}{\hbar}\hat{H}_{S}(\tau-t)}=\sum^{\infty}_{n=0}\frac{1}{n!}\frac{i^{n}}{\hbar^{n}}\left(t-\tau\right)^{n}\left[\hat{f}_{S},\hat{H}_{S}\right]_{(n)}
\end{equation}
where
\begin{equation}
\left[\hat{f}_{S},\hat{H}_{S}\right]_{(n)}=\left[\left[\hat{f}_{S},\hat{H}_{S}\right]_{(n-1)},\hat{H}_{S}\right],\left[\hat{f}_{S},\hat{H}_{S}\right]_{(0)}\coloneqq\hat{f}_{S}.\label{eq:commutator-n}
\end{equation}
Then the gauge-invariant relational observable associated to $\hat{f}_{S}$
is derived by projecting the above into the clock states \cite{H_hn_2021,Dittrich:2004cb,Dittrich_2006,Rovelli:2011fk}
\begin{equation}
\hat{F}_{f_{S},T}\left(\tau\right)=\frac{1}{2\pi\hbar}\int^{+\infty}_{-\infty}dt|t\rangle_{g}\langle t|_{g}\sum^{\infty}_{n=0}\frac{1}{n!}\frac{i^{n}}{\hbar^{n}}\left(t-\tau\right)^{n}\left[\hat{f}_{S},\hat{H}_{S}\right]_{(n)},\label{eq:gauge-inv-obs-general-Sum}
\end{equation}
or equivalently
\begin{equation}
\begin{aligned}\hat{F}_{f_{S},T}\left(\tau\right)= & \frac{1}{2\pi\hbar}\int^{+\infty}_{-\infty}dt|t\rangle_{g}\langle t|_{g}e^{\frac{i}{\hbar}(\tau-t)\hat{H}_{S}}\hat{f}_{S}e^{-\frac{i}{\hbar}(\tau-t)\hat{H}_{S}}\\
= & \frac{1}{2\pi\hbar}\int^{+\infty}_{-\infty}dte^{-\frac{i}{\hbar}t\left(\hat{H}_{C}+\hat{H}_{S}\right)}\left(|\tau\rangle_{g}\langle\tau|_{g}\hat{f}_{S}\right)e^{\frac{i}{\hbar}t\left(\hat{H}_{C}+\hat{H}_{S}\right)}\\
= & \frac{1}{2\pi}\int_{R}ds\,\hat{U}_{CS}(s)\left(|\tau\rangle_{g}\langle\tau|_{g}\hat{f}_{S}\right)\hat{U}^{\dagger}_{CS}(s)\\
= & \eta\left(|\tau\rangle_{g}\langle\tau|_{g}\hat{f}_{S}\right),
\end{aligned}
\label{eq:gauge-inv-obs-general}
\end{equation}
where 
\begin{equation}
\hat{U}_{CS}(s)=\exp\left[-is\hat{C}_{H}\right],\label{eq:UCS}
\end{equation}
and in our case
\begin{equation}
\hat{C}_{H}=\alpha^{2}\widehat{\frac{1}{p_{a}}}+\hat{p}_{b}.\label{eq:CH-quantum}
\end{equation}
Eqs. \eqref{eq:gauge-inv-obs-general} or \eqref{eq:gauge-inv-obs-general-Sum}
yields the gauge-invariant value associated to $\hat{f}_{S}$ when
the clock operator $\hat{T}$ has eigenvalue $\tau$.

The action of $\hat{F}_{f_{S},T}\left(\tau\right)$ on a physical
state $\left|\Psi^{\text{phys}}_{\psi}\right\rangle $ for which $\hat{U}^{\dagger}_{CS}\left|\Psi^{\text{phys}}_{\psi}\right\rangle =\left|\Psi^{\text{phys}}_{\psi}\right\rangle $
yields
\begin{equation}
\begin{aligned}\hat{F}_{f_{S},T}\left(\tau\right)\left|\Psi^{\text{phys}}_{\psi}\right\rangle = & \frac{1}{2\pi}\int_{\mathbb{R}}ds\,\hat{U}_{CS}\left(s\right)\left\{ |\tau\rangle_{g}\langle\tau|_{g}\hat{f}_{S}\right\} \hat{U}^{\dagger}_{CS}\left(s\right)\left|\Psi^{\text{phys}}_{\psi}\right\rangle \\
= & \hat{\Pi}_{\text{phys}}\left\{ |\tau\rangle_{g}\langle\tau|_{g}\hat{f}_{S}\right\} \left|\Psi^{\text{phys}}_{\psi}\right\rangle ,
\end{aligned}
\end{equation}
where the projector into the physical states $\hat{\Pi}_{\text{phys}}:\Phi\to\Phi^{*}$
is defined as 
\begin{equation}
\hat{\Pi}_{\text{phys}}=\frac{1}{2\pi}\int_{\mathbb{R}}ds\,\hat{U}_{CS}\left(s\right)=\hat{\Pi}^{\dagger}_{\text{phys}}.\label{eq:Pi-phys-projector}
\end{equation}
The hermiticity of $\hat{\Pi}_{\text{phys}}$ can be checked easily
given the integral is over $\mathbb{R}$. Hence, Eq. \eqref{eq:gauge-inv-obs-general}
can equivalently be written as
\begin{equation}
\hat{F}_{f_{S},T}\left(\tau\right)=\eta\left(|\tau\rangle_{g}\langle\tau|_{g}\hat{f}_{S}\right)\approx\hat{\Pi}_{\text{phys}}\left\{ |\tau\rangle_{g}\langle\tau|_{g}\hat{f}_{S}\right\} ,
\end{equation}
where $\approx$ means that the above equality only holds for physical
states. As a result the physical expectation value of $\hat{F}_{f_{S},T}\left(\tau\right)$
can be written as
\begin{equation}
\begin{aligned}\left\langle \Psi^{\text{phys}}_{\psi}\left|\hat{F}_{f_{S},T}\left(\tau\right)\right|\Psi^{\text{phys}}_{\psi}\right\rangle _{\text{phys}}= & \left\langle \psi\left|\hat{\Pi}_{\text{phys}}\left\{ \left|\tau\right\rangle _{g}\left\langle \tau\right|_{g}\hat{f}_{S}\right\} \right|\Psi^{\text{phys}}_{\psi}\right\rangle _{\mathrm{kin}}\\
= & \left\langle \psi\left|\hat{\Pi}_{\text{phys}}\right|\tau_{g}\right\rangle _{\text{kin}}\left\langle \tau_{g}\left|\hat{f}_{S}\right|\Psi^{\text{phys}}_{\psi}\right\rangle _{\text{kin}}\\
= & \left\langle \psi\left|\hat{\Pi}^{\dagger}_{\text{phys}}\right|\tau_{g}\right\rangle _{\text{kin}}\left\langle \tau_{g}\left|\hat{f}_{S}\right|\Psi^{\text{phys}}_{\psi}\right\rangle _{\text{kin}}\\
= & \left\langle \Psi^{\text{phys}}_{\psi}\bigg|\tau_{g}\right\rangle \left\langle \tau_{g}\left|\hat{f}_{S}\right|\Psi^{\text{phys}}_{\psi}\right\rangle _{\text{kin}}\\
= & \left\langle \Psi_{\text{PW}}\left(\tau_{g}\right)\right|\hat{f}_{S}\left|\Psi_{\text{PW}}\left(\tau_{g}\right)\right\rangle 
\end{aligned}
\label{eq:F-Ab-phys-phys}
\end{equation}
where the last line is derived assuming $\hat{f}_{S}$ only includes
system operators, and we have identified the Page-Wootters (PW) states
as 
\begin{equation}
\left|\Psi_{\text{PW}}\left(\tau_{g}\right)\right\rangle =\left\langle \tau_{g}\bigg|\Psi^{\text{phys}}_{\psi}\right\rangle .
\end{equation}
This shows that $\mathscr{H}_{\text{phys}}$ and the PW reduced Hilbert
space $\mathscr{H}_{\text{PW}}$ are isometric as was pointed out
in \cite{H_hn_2021}. Consequently, the PW scalar product and matrix
elements are gauge-invariant too. In our system, we can compute the
terms in \eqref{eq:F-Ab-phys-phys} more explicitly, using Eqs. \eqref{eq:physical-states-ket}
and \eqref{eq:Fourier-b-t-g-} as
\begin{equation}
\begin{aligned}\left\langle \tau_{g}\bigg|\Psi^{\text{phys}}_{\psi}\right\rangle = & \left\langle \tau\right|_{g}\int_{\mathbb{R}}dp_{b}\int_{\mathbb{R}}dp_{a}\,\psi_{a|b}\left(p_{a}\right)\delta\left(p_{b}+\frac{\alpha^{2}}{p_{a}}\right)\left|p_{a}\right\rangle _{a}\left|p_{b}\right\rangle _{b}\\
= & \int_{\mathbb{R}}dp_{b}\int_{\mathbb{R}}dp_{a}\,\psi_{a|b}\left(p_{a}\right)\delta\left(p_{b}+\frac{\alpha^{2}}{p_{a}}\right)e^{-ig\left(p_{b}\right)}e^{\frac{i}{\hbar}p_{b}\tau}\left|p_{a}\right\rangle _{a},
\end{aligned}
\label{eq:PW-states}
\end{equation}
For the matrix element $\left\langle \tau_{g}\left|\hat{f}_{S}\right|\Psi^{\text{phys}}_{\psi}\right\rangle _{\text{kin}}$,
if in general $\hat{f}_{S}\left(\hat{a},\hat{p}_{a},\hat{b},\hat{p}_{b}\right)$,
then we obtain
\begin{equation}
\begin{aligned}\left\langle \tau_{g}\left|\hat{f}_{S}\right|\Psi^{\text{phys}}_{\psi}\right\rangle _{\text{kin}}= & \left\langle \tau\right|_{g}\hat{f}_{S}\int_{\mathbb{R}}dp_{b}\int_{\mathbb{R}}dp_{a}\,\psi_{a|b}\left(p_{a}\right)\delta\left(p_{b}+\frac{\alpha^{2}}{p_{a}}\right)\left|p_{a}\right\rangle _{a}\left|p_{b}\right\rangle _{b}\\
= & \int_{\mathbb{R}}dp^{\prime}_{b}\int_{\mathbb{R}}dp_{b}\int_{\mathbb{R}}dp_{a}\,\psi_{a|b}\left(p_{a}\right)\delta\left(p_{b}+\frac{\alpha^{2}}{p_{a}}\right)e^{-ig\left(p^{\prime}_{b}\right)}e^{\frac{i}{\hbar}p^{\prime}_{b}\tau}\\
 & \times{}_{b}\left\langle p^{\prime}_{b}\right|\hat{f}_{S}\left|p_{b}\right\rangle _{b}\left|p_{a}\right\rangle _{a}
\end{aligned}
\label{eq:PW-tau-f-state-general}
\end{equation}
Notice that $_{b}\left\langle p^{\prime}_{b}\right|\hat{f}_{S}\left|p_{b}\right\rangle _{b}$
would still be an operator if $f$ depends on $a,\,p_{a}$. In the
simplest case that $\hat{f}_{S}=\hat{f}_{S}\left(\hat{a},\hat{p}_{a}\right)$
this reduces to
\begin{equation}
\left\langle \tau_{g}\left|\hat{f}_{S}\right|\Psi^{\text{phys}}_{\psi}\right\rangle _{\text{kin}}=\int_{\mathbb{R}}dp_{b}\int_{\mathbb{R}}dp_{a}\,\psi_{a|b}\left(p_{a}\right)\delta\left(p_{b}+\frac{\alpha^{2}}{p_{a}}\right)e^{-ig\left(p_{b}\right)}e^{\frac{i}{\hbar}p_{b}\tau}\hat{f}_{S}\left|p_{a}\right\rangle _{a}\label{eq:PW-tau-f-state-f-only-a-pa}
\end{equation}
Finally replacing \eqref{eq:PW-states} and \eqref{eq:PW-tau-f-state-general}
in \eqref{eq:F-Ab-phys-phys} we obtain
\begin{equation}
\begin{aligned}\left\langle \Psi^{\text{phys}}_{\psi}\left|\hat{F}_{f_{S},T}\left(\tau\right)\right|\Psi^{\text{phys}}_{\psi}\right\rangle _{\text{phys}}= & \int_{\mathbb{R}}dp^{\prime\prime}_{b}\int_{\mathbb{R}}dp^{\prime\prime}_{a}\int_{\mathbb{R}}dp^{\prime}_{b}\int_{\mathbb{R}}dp_{b}\int_{\mathbb{R}}dp_{a}\,\\
 & \times\psi^{*}_{a|b}\left(p^{\prime\prime}_{a}\right)\psi_{a|b}\left(p_{a}\right)\delta\left(p^{\prime\prime}_{b}+\frac{\alpha^{2}}{p^{\prime\prime}_{a}}\right)\delta\left(p_{b}+\frac{\alpha^{2}}{p_{a}}\right)\\
 & \times e^{-i\left(g\left(p^{\prime}_{b}\right)-g\left(p^{\prime\prime}_{b}\right)\right)}e^{\frac{i}{\hbar}\left(p^{\prime}_{b}-p^{\prime\prime}_{b}\right)\tau}\\
 & \times{}_{a}\left\langle p^{\prime\prime}_{a}\right|\left(_{b}\left\langle p^{\prime}_{b}\right|\hat{f}_{S}\left|p_{b}\right\rangle _{b}\right)\left|p_{a}\right\rangle _{a}
\end{aligned}
\label{eq:expect-gauge-inv-obs-general-g}
\end{equation}
Again if $\hat{f}_{S}$ depends only on $a,\,p_{a}$, then the above
reduces to 
\begin{equation}
\begin{aligned}\left\langle \Psi^{\text{phys}}_{\psi}\left|\hat{F}_{f_{S},T}\left(\tau\right)\right|\Psi^{\text{phys}}_{\psi}\right\rangle _{\text{phys}}= & \int_{\mathbb{R}}dp^{\prime}_{a}\int_{\mathbb{R}}dp_{a}\,\psi^{*}_{a|b}\left(p^{\prime}_{a}\right)\psi_{a|b}\left(p_{a}\right)\\
 & \times e^{\frac{i}{\hbar}\left(\tilde{g}\left(p_{a}\right)-\tilde{g}\left(p^{\prime}_{a}\right)\right)}e^{-\frac{i\alpha^{2}}{\hbar}\left(\frac{1}{p_{a}}-\frac{1}{p^{\prime}_{a}}\right)\tau}{}_{a}\left\langle p^{\prime}_{a}\right|\hat{f}_{S}\left|p_{a}\right\rangle _{a},
\end{aligned}
\label{eq:expect-gauge-inv-obs-f-a-pa-g}
\end{equation}
where we have defined
\[
\tilde{g}\left(p_{a}\right)\coloneqq-\hbar g\left(-\frac{\alpha^{2}}{p_{a}}\right).
\]
Note that if $\hat{f}_{S}$ depends on clock DoF $(b,\,p_{b})$ then
from \eqref{eq:gauge-inv-obs-general} we get 
\begin{equation}
\begin{aligned}\hat{F}^{\dagger}_{f_{S},T}\left(\tau\right)= & \frac{1}{2\pi}\int_{R}ds\,\hat{U}_{CS}(s)\left(\hat{f}^{\dagger}_{S}|\tau\rangle_{g}\langle\tau|_{g}\right)\hat{U}^{\dagger}_{CS}(s)\end{aligned}
\end{equation}
Hence, assuming a symmetric operator such that $\hat{f}_{S}=\hat{f}^{\dagger}_{S}$,
then 
\begin{equation}
\hat{F}_{f_{S},T}\left(\tau\right)=\hat{F}^{\dagger}_{f_{S},T}\left(\tau\right)\Leftrightarrow\left[\hat{f}_{S},|\tau\rangle_{g}\langle\tau|_{g}\right]=0.\label{eq:Reality-Condition}
\end{equation}
Since from \eqref{eq:Fourier-b-t-g-} $|\tau\rangle_{g}$ can be written
as 
\begin{equation}
|\tau\rangle_{g}=\int_{\mathbb{R}}dp_{b}\,e^{ig\left(p_{b}\right)}e^{-\frac{i}{\hbar}p_{b}\tau}\left|p_{b}\right\rangle _{b},
\end{equation}
the condition \eqref{eq:Reality-Condition} holds for $\hat{f}_{S}\left(a,p_{a}\right)$.
However, an operator $\hat{f}_{S}$ that also depends on $\hat{b}$
or $\hat{p}_{b}$ or both will not fulfill the requirement \eqref{eq:Reality-Condition},
e.g.,
\begin{equation}
\hat{p}_{b}|\tau\rangle_{g}=\int_{\mathbb{R}}dp_{b}\,e^{ig\left(p_{b}\right)}e^{-\frac{i}{\hbar}p_{b}\tau}p_{b}\left|p_{b}\right\rangle _{b}\neq\tau|\tau\rangle_{g}
\end{equation}
and one should use
\begin{equation}
\begin{aligned}\left\langle \Psi^{\text{phys}}_{\psi}\left|\frac{1}{2}\left(\hat{F}_{f_{S},T}\left(\tau\right)+\hat{F}^{\dagger}_{f_{S},T}\left(\tau\right)\right)\right|\Psi^{\text{phys}}_{\psi}\right\rangle _{\text{phys}}= & \frac{1}{2}\left\langle \Psi^{\text{phys}}_{\psi}\left|\hat{F}_{f_{S},T}\left(\tau\right)\right|\Psi^{\text{phys}}_{\psi}\right\rangle _{\text{phys}}\\
 & +\frac{1}{2}\left(\left\langle \Psi^{\text{phys}}_{\psi}\left|\hat{F}_{f_{S},T}\left(\tau\right)\right|\Psi^{\text{phys}}_{\psi}\right\rangle _{\text{phys}}\right)^{*}\\
= & \mathfrak{Re}\left(\left\langle \Psi^{\text{phys}}_{\psi}\left|\hat{F}_{f_{S},T}\left(\tau\right)\right|\Psi^{\text{phys}}_{\psi}\right\rangle _{\text{phys}}\right)
\end{aligned}
\label{eq:Re-expectation-general}
\end{equation}
to compute the expectation value of the gauge-invaraint extension
of $\hat{f}_{S}$.

We will use Eqs. \eqref{eq:expect-gauge-inv-obs-general-g}, \eqref{eq:expect-gauge-inv-obs-f-a-pa-g}
and \eqref{eq:Re-expectation-general} for most of the computations
in the next sections to study singularity related physical quantities.

\subsection{Area of 2-spheres}\label{sec:area-evol}

In a Schwarzschild black hole, the area of two sphere $A$ is related
to the existence of a physical singularity via the Kretschmann scalar
$K\propto\frac{1}{r^{6}}\propto\frac{1}{A^{3}}$. A vanishing $A$
thus means that the Kretschmann blows up \cite{Chiou_2008,Fragomeno_2025}.
Therefore, we would like to study the behavior of the quantum version
of this area under the evolution in the relation approach as a means
of studying the singularity. Given the classical expression of $A$
in \eqref{eq:Area-a-class}, one might be tempted to define the area
operator as 
\begin{equation}
\hat{f}_{S}=\hat{A}=4\pi\hat{a}^{2}.\label{eq:A-gauge-depend}
\end{equation}
However, such an operator is not gauge-invariant since 
\begin{equation}
\left[\hat{a}^{2},\hat{C}_{H}\right]=\left[\hat{a}^{2},\widehat{\tfrac{1}{p_{a}}}\right]\neq0
\end{equation}
and thus cannot correspond to a physical observable. Although not
necessary for our computations of the expectation value of the gauge-invariant
area operator,
we can construct the gauge-invariant extension of $\hat{A}$ itself
using \eqref{eq:gauge-inv-obs-general-Sum} and \eqref{eq:commutator-n}. To find an explicit expression for the gauge-invariant area operator,
we first note that
\begin{equation}
\left[\hat{A},\widehat{\frac{1}{p_{a}}}\right]_{(0)}:=\hat{A}.
\end{equation}
To compute the next $n=1$ order commutator in Eq. \eqref{eq:commutator-n},
and assuming that $\widehat{\tfrac{1}{p_{a}}}$ is well-defined on
the Hilbert space (which it is due to our discussion of the previous
section), we can find
\begin{equation}
0=\left[\hat{a},\hat{\mathbb{I}}\right]=\left[\hat{a},\hat{p}_{a}\widehat{\tfrac{1}{p_{a}}}\right]=i\hbar\widehat{\tfrac{1}{p_{a}}}+\hat{p}_{a}\left[\hat{a},\widehat{\tfrac{1}{p_{a}}}\right]
\end{equation}
and thus
\begin{equation}
\left[\hat{a},\widehat{\tfrac{1}{p_{a}}}\right]=-i\hbar\left(\widehat{\tfrac{1}{p_{a}}}\right)^{2}.\label{eq:a-p-1-commut}
\end{equation}
As a result one obtains
\begin{equation}
\begin{aligned}\left[\hat{A},\alpha^{2}\widehat{\frac{1}{p_{a}}}\right]_{(1)}= & -4\pi\alpha^{2}i\hbar\left\{ \hat{a}\left(\widehat{\tfrac{1}{p_{a}}}\right)^{2}+\left(\widehat{\tfrac{1}{p_{a}}}\right)^{2}\hat{a}\right\} ,\\
\left[\hat{A},\alpha^{2}\widehat{\frac{1}{p_{a}}}\right]_{(2)}= & -8\pi\hbar^{2}\alpha^{4}\left(\widehat{\tfrac{1}{p_{a}}}\right)^{4},\\
\left[\hat{A},\alpha^{2}\widehat{\frac{1}{p_{a}}}\right]_{(k)}= & 0,\qquad\text{for }k\geq3,
\end{aligned}
\end{equation}
Plugging these expressions into Eq. \eqref{eq:gauge-inv-obs-general-Sum}
yields
\begin{equation}
\begin{aligned}\hat{F}_{A,b}\left(\tau\right)= & 4\pi\left(\hat{a}^{2}+\alpha^{2}\left(\hat{b}-\tau\right)\left\{ \hat{a}\left(\widehat{\tfrac{1}{p_{a}}}\right)^{2}+\left(\widehat{\tfrac{1}{p_{a}}}\right)^{2}\hat{a}\right\} +\alpha^{4}\left(\hat{b}-\tau\right)^{2}\left(\widehat{\tfrac{1}{p_{a}}}\right)^{4}\right)\\
= & 4\pi\left(\hat{a}+\alpha^{2}\left(\hat{b}-\tau\right)\left(\widehat{\tfrac{1}{p_{a}}}\right)^{2}\right)^{2}\\
= & 4\pi\hat{a}^{2}_{\text{GI}}\left(\tau\right),
\end{aligned}
\label{eq:gauge-inv-area-operator}
\end{equation}
where we have defined the gauge-invariant extension of $a$ as
\begin{equation}
\hat{a}_{\text{GI}}\left(\tau\right)=\hat{a}+\alpha^{2}\left(\hat{b}-\tau\right)\left(\widehat{\tfrac{1}{p_{a}}}\right)^{2}.
\end{equation}
Interestingly the gauge invariant version of the area operator includes
two correction terms linear and quadratic in $(\hat{b}-\tau)$. As a result,
such a gauge invariant area is a parabola as a function of the relational
time $b$, which clearly will have a minimum, and allows the possibility
of a bounce scenario.

Given that $\hat{a}^2$ only depends on $a$, the expectation value
of its gauge-invariant extension can be written, using \eqref{eq:expect-gauge-inv-obs-f-a-pa-g},
as
\begin{equation}
\begin{aligned}\left\langle \Psi^{\text{phys}}_{\psi}\left|\hat{F}_{A,T}\left(\tau\right)\right|\Psi^{\text{phys}}_{\psi}\right\rangle _{\text{phys}}= & 4\pi\int_{\mathbb{R}}dp^{\prime}_{a}\int_{\mathbb{R}}dp_{a}\,\psi^{*}_{a|b}\left(p^{\prime}_{a}\right)\psi_{a|b}\left(p_{a}\right)\\
 & \times e^{-\frac{i\alpha^{2}}{\hbar}\tau\left(\frac{1}{p_{a}}-\frac{1}{p^{\prime}_{a}}\right)}e^{\frac{i}{\hbar}\left(\tilde{g}\left(p_{a}\right)-\tilde{g}\left(p^{\prime}_{a}\right)\right)}\left\langle p^{\prime}_{a}\left|\hat{a}^{2}\right|p_{a}\right\rangle _{a}.
\end{aligned}
\end{equation}
Using 
\begin{equation}
\left\langle p^{\prime}_{a}\left|\hat{a}^{2}\right|p_{a}\right\rangle _{a}=-\hbar^{2}\frac{\partial^{2}}{\partial p^{2}_{a}}\delta\left(p^{\prime}_{a}-p_{a}\right)
\end{equation}
and assuming the functions involved decay fast enough at infinities,
we get
\begin{equation}
\begin{aligned}\left\langle \hat{F}_{A,T}\left(\tau\right)\right\rangle _{\text{phys}}= & 4\pi\int_{\mathbb{R}}dp_{a}\,\left(\left(\frac{\partial\tilde{g}\left(p_{a}\right)}{\partial p_{a}}+\frac{\alpha^{2}}{p^{2}_{a}}\tau\right)^{2}\left|\psi_{a|b}\left(p_{a}\right)\right|^{2}+\hbar^{2}\left|\frac{\partial\psi_{a|b}\left(p_{a}\right)}{\partial p_{a}}\right|^{2}\right.\\
 & \left.+i\hbar\left(\frac{\partial\tilde{g}\left(p_{a}\right)}{\partial p_{a}}+\frac{\alpha^{2}}{p^{2}_{a}}\tau\right)\left[\frac{\partial\psi^{*}_{a|b}\left(p_{a}\right)}{\partial p_{a}}\psi_{a|b}\left(p_{a}\right)-\psi^{*}_{a|b}\left(p_{a}\right)\frac{\partial\psi_{a|b}\left(p_{a}\right)}{\partial p_{a}}\right]\right)\\
= & 4\pi\int_{\mathbb{R}}dp_{a}\,\left|\left(\frac{\partial\tilde{g}\left(p_{a}\right)}{\partial p_{a}}+\frac{\alpha^{2}}{p^{2}_{a}}\tau\right)\psi_{a|b}\left(p_{a}\right)-i\hbar\frac{\partial\psi_{a|b}\left(p_{a}\right)}{\partial p_{a}}\right|^{2},
\end{aligned}
\label{eq:F-Ab-phys-phys-explicit}
\end{equation}
where the function $g\left(p_{b}\right)$ must be smooth and its derivative
must not grow too fast as $p_{a}\rightarrow0$ (i.e., as $p_{b}\rightarrow\infty$
). We see that the second line above mimics the probability current
density in quantum mechanics. More importantly, the area expectation
value has a parabolic dependence on $\tau$ as $\left\langle \hat{F}_{A,b}\left(\tau\right)\right\rangle _{\text{phys}}=A\tau^{2}+B\tau+C$.
This can be interpreted as contraction of 2-spheres in the trapped
black hole region, reaching to a minimum area, followed by a bounce
to the anti-trapped white hole spacetime, resulting in a black hole-to-white
hole transition picture.

The result \eqref{eq:F-Ab-phys-phys-explicit} shows that the gauge-invariant
area is proportional to the integral of a modulus squared function,
which is nonnegative as it should be. Below we show that it is also nonzero which makes it strictly positive. Let us for a moment consider
the case where $g=0$, i.e.,
\begin{equation}
\left\langle \Psi^{\text{phys}}_{\psi}\left|\hat{F}_{A,b}\left(\tau\right)\right|\Psi^{\text{phys}}_{\psi}\right\rangle _{\text{phys}}=4\pi\int_{\mathbb{R}}dp_{a}\,\left|\frac{\alpha^{2}}{p^{2}_{a}}\psi_{a|b}\left(p_{a}\right)\tau-i\hbar\frac{d\psi_{a|b}\left(p_{a}\right)}{dp_{a}}\right|^{2}
\end{equation}
This will only vanish if for some $\tau=\tau_{0}$, the integrand
vanishes. This would correspond to the classical singularity. The
vanishing of this integrand yields a differential equation
\begin{equation}
\frac{d\psi_{a|b}\left(p_{a}\right)}{dp_{a}}=-\frac{i}{\hbar}\frac{\alpha^{2}\tau}{p^{2}_{a}}\psi_{a|b}\left(p_{a}\right)\label{eq:ODE-zero-area}
\end{equation}
 with a solution
\begin{equation}
\psi_{a|b}\left(p_{a};\tau_{0}\right)=Ce^{\frac{i}{\hbar}\frac{\alpha^{2}}{p_{a}}\tau_{0}},\label{eq:psi-ab-zero-area}
\end{equation}
with $C$ being an integration constant. But such a $\psi_{a|b}$
is non-normalizable and hence does not belong to $\mathscr{H}_{\text{phys}}$.
One might try to construct a normalizable state out of the above functions
as
\begin{equation}
\varUpsilon\left(p_{a}\right)=\int d\tau^{\prime}\,\xi\left(\tau^{\prime}\right)e^{\frac{i\alpha^{2}}{\hbar p_{a}}\tau^{\prime}}.\label{eq:psi-xi-zero-area}
\end{equation}
Plugging this in \eqref{eq:ODE-zero-area} leads to the condition
$(\tau^{\prime}-\tau_{0})\xi(\tau^{\prime})=0,\,\forall\tau^{\prime}$.
The only nontrivial function satisfying this condition is $\xi(\tau')=C\delta(\tau'-\tau_{0})$.
Plugging this back into \eqref{eq:psi-xi-zero-area} again yields
\eqref{eq:psi-ab-zero-area}. We thus conclude that there is no physical
state or solution for which $\left\langle \hat{F}_{A,b}\left(\tau\right)\right\rangle _{\text{phys}}=0$. This result holds not only for the special case in which $T=b$, but it is valid for any clock \eqref{eq:T-Tg0}: the expectation value \eqref{eq:F-Ab-phys-phys-explicit} will vanish only for the non-normalizable states
\begin{equation}    
\psi_{a|b}\left(p_{a};\tau_{0}\right)= C e^{\frac{i}{\hbar}\big( \frac{\alpha^2\tau_0}{p_a}-\tilde{g}(p_a) \big)}
\end{equation}
for a generic function $\tilde{g}\left(p_{a}\right)\coloneqq-\hbar g\left(-\frac{\alpha^{2}}{p_{a}}\right)$. This means that the vanishing of the expectation value of the area of 2-spheres in this model is valid for any choice of a quantum clock \eqref{eq:T-Tg0}.

This is a strong
hint of singularity resolution, although not a direct proof. Hence
in what follows, we prove this statement by studying the gauge-invariant
extensions of the Kretschmann scalar and the expansion scalar. 

\subsection{Expansion scalar}\label{sec:expansion-scalar}

The expansion scalar $\vartheta$, particularly for null geodesics,
is one of the most important indicators of whether or not a spacetime
is singular. It describes whether the geodesics during their motion
through spacetime get focused or defocused. An infinite focusing, corresponding
to $\vartheta\to-\infty$ in some region of spacetime, signals the
presence of singularity in that region. 

Given a congruence of null geodesics described by a null tangent vector
field $k^{\mu}$, the expansion scalar for such a congruence that
is affinely parametrized, i.e., 
\begin{equation}
k^{\mu}\nabla_{\mu}k^{\nu}=0,\qquad k_{\mu}k^{\mu}=0
\end{equation}
can be computed as
\begin{equation}
\vartheta=\nabla_{\mu}k^{\mu}
\end{equation}
where $\nabla_{\mu}$ is the covariant derivative associated to the
Levi-Civita connection of the spacetime.

For a radial null geodesic
\begin{equation}
k^{\mu}=\left(k^{0}\left(x,\lambda\right),k^{1}\left(x,\lambda\right),0,0\right)
\end{equation}
in the spacetime described by the metric \eqref{eq:metric-ab} we obtain
\begin{equation}
\vartheta= \frac{1}{N} \frac{\dot{A}}{A} =-\frac{2}{\alpha}\frac{p_{b}}{a},
\end{equation}
where $ \dot{A}:=\frac{d}{d \lambda}\left( 4 \pi a(\lambda)^2  \right) $. This classical gauge-dependent expression is different from
the expression of the area of 2-spheres in that it depends on clock
DoF $p_{b}$. As discussed in \ref{subsec:General-treatment-obs},
we should consider the real part of its expectation value.

The gauge-dependent expansion scalar operator is
\begin{equation}
\hat{f}_{S}=\hat{\vartheta}=-\frac{2}{\alpha}\hat{p}_{b}\widehat{\frac{1}{a}}.\label{eq:theta-in-theta-tilde}
\end{equation}
The expectation value of the gauge-invariant extension of this operator
is derived using \eqref{eq:expect-gauge-inv-obs-general-g} as
\begin{equation}
\begin{aligned}\mathfrak{Re}\left\langle \hat{F}_{\vartheta,T}\left(\tau\right)\right\rangle _{\text{phys}}= & \mathfrak{Re}\left\{ -\frac{2}{\alpha}\int_{\mathbb{R}}dp^{\prime}_{b}\int_{\mathbb{R}}dp^{\prime}_{a}\int_{\mathbb{R}}dp_{b}\int_{\mathbb{R}}dp_{a}\right.\\
 & \times p_{b}\psi^{*}_{a|b}\left(p^{\prime}_{a}\right)\psi_{a|b}\left(p_{a}\right)\delta\left(p^{\prime}_{b}+\frac{\alpha^{2}}{p^{\prime}_{a}}\right)\delta\left(p_{b}+\frac{\alpha^{2}}{p_{a}}\right)\\
 & \times e^{-i\left(g\left(p_{b}\right)-g\left(p^{\prime}_{b}\right)\right)}e^{\frac{i}{\hbar}\left(p_{b}-p^{\prime}_{b}\right)\tau}\\
 & \left.\times{}_{a}\left\langle p^{\prime}_{a}\right|\widehat{\frac{1}{a}}\left|p_{a}\right\rangle _{a}\right\} 
\end{aligned}
\label{eq:theta-intermediate}
\end{equation}
To derive a general expression for $\left\langle p^{\prime}_{a}\left|\widehat{\frac{1}{a^{n}}}\right|p_{a}\right\rangle $
we consider 
\begin{equation}
\hat{a}^{n}\frac{\widehat{1}}{a^{n}}=\mathbb{I},
\end{equation}
and compute its matrix element as
\begin{equation}
\left\langle p^{\prime}_{a}\left|\hat{a}^{n}\frac{\widehat{1}}{a^{n}}\right|p_{a}\right\rangle =\delta\left(p^{\prime}_{a}-p_{a}\right).
\end{equation}
Using $\left\langle p^{\prime}_{a}\right|\hat{a}=i\hbar\frac{\partial}{\partial p^{\prime}_{a}}\left\langle p^{\prime}_{a}\right|$
yields
\begin{equation}
(i\hbar)^{n}\frac{\partial^{n}}{\partial p^{\prime n}_{a}}\left\langle p^{\prime}_{a}\left|\frac{\widehat{1}}{a^{n}}\right|p_{a}\right\rangle =\delta\left(p^{\prime}_{a}-p_{a}\right)
\end{equation}
which is an $n$-th order differential equation for the Green's function
$G_{n}\left(p^{\prime}_{a}-p_{a}\right)$. The solution to this equation
is
\begin{equation}
\begin{aligned}\left\langle p^{\prime}_{a}\left|\frac{\widehat{1}}{a^{n}}\right|p_{a}\right\rangle = & \left(\frac{1}{i\hbar}\right)^{n}G_{n}\left(p^{\prime}_{a}-p_{a}\right)\\
= & \left(\frac{1}{i\hbar}\right)^{n}\frac{\left(p^{\prime}_{a}-p_{a}\right)^{n-1}\text{sgn}\left(p^{\prime}_{a}-p_{a}\right)}{2\left(n-1\right)!}\\
= & \left(\frac{1}{i\hbar}\right)^{n}\frac{\left|p^{\prime}_{a}-p_{a}\right|^{n-1}\left[\text{sgn}\left(p^{\prime}_{a}-p_{a}\right)\right]^{n}}{2\left(n-1\right)!}.
\end{aligned}
\label{eq:Green-inverse-a}
\end{equation}
 Notice that from this computation, there is no mathematical condition
that exclude states with non-null support at $a=0$. Also, unlike
the exclusion of states with $p_{a}=0$, which ensures the monotonicity
of the clock, the $a=0$ exclusion is not supported by any physical
reason, although it could instead force the singularity resolution
by simply not considering the cases in which $a=0$. Thus the black
hole is allowed to have states with support in $a=0$.

Plugging in the \eqref{eq:Green-inverse-a} into yields
\begin{equation}
\begin{aligned}\mathfrak{Re}\left\langle \Psi^{\text{phys}}_{\psi}\left|\hat{F}_{\vartheta,T}\left(\tau\right)\right|\Psi^{\text{phys}}_{\psi}\right\rangle _{\text{phys}}= & \mathfrak{Re}\left\{ \frac{i\alpha}{\hbar}\int_{\mathbb{R}}dp^{\prime}_{a}\int_{\mathbb{R}}dp_{a}\,\psi^{*}_{a|b}\left(p^{\prime}_{a}\right)\frac{1}{p_{a}}\psi_{a|b}\left(p_{a}\right)\right.\\
 & \left.\times e^{\frac{i}{\hbar}\left(\tilde{g}\left(p_{a}\right)-\tilde{g}\left(p^{\prime}_{a}\right)\right)}e^{\frac{i}{\hbar}\left(-\frac{\alpha^{2}}{p_{a}}+\frac{\alpha^{2}}{p^{\prime}_{a}}\right)\tau}\text{sgn}\left(p_{a}-p^{\prime}_{a}\right)\right\} 
\end{aligned}
\label{eq:F-theta-phys-exp-fin}
\end{equation}
In the classical regime, a singularity is present whenever $\vartheta\to-\infty$.
As a result we want to check whether in the quantum relational approach,
we still have such behavior, i.e., whether $\mathfrak{Re}\left\langle \hat{F}_{\vartheta,T}\left(\tau\right)\right\rangle _{\text{phys}}\to-\infty$
for some $\tau$. To this end we note that
\begin{equation}
\left|\mathfrak{Re}\left\langle \hat{F}_{f_{S},T}\left(\tau\right)\right\rangle _{\text{phys}}\right|\leq\left|\left\langle \hat{F}_{f_{S},T}\left(\tau\right)\right\rangle _{\text{phys}}\right|=\left|\int dp^{\prime}_{a}dp_{a}\ldots\right|\leq\int dp^{\prime}_{a}dp_{a}\left|\ldots\right|\label{eq:Abs-int-condition}
\end{equation}
Applying this to \eqref{eq:F-theta-phys-exp-fin} yields
\begin{equation}
\left|\mathfrak{Re}\left\langle \Psi^{\text{phys}}_{\psi}\left|\hat{F}_{\vartheta,T}\left(\tau\right)\right|\Psi^{\text{phys}}_{\psi}\right\rangle _{\text{phys}}\right|\leq\frac{\alpha}{\hbar}\int_{\mathbb{R}}dp^{\prime}_{a}\int_{\mathbb{R}}dp_{a}\,\left|\psi^{*}_{a|b}\left(p^{\prime}_{a}\right)\right|\frac{1}{\left|p_{a}\right|}\left|\psi_{a|b}\left(p_{a}\right)\right|<\infty.
\end{equation}
In obtaining this result, i.e., non-divergence of the integral, we
have used the properties of the Thiemann states \eqref{eq:physical-state-Thiemann}
and the physical Hilbert space $\mathscr{H}_{\text{phys}}$ as discussed
before. In other words, since $\psi_{a\mid b}\in D_{0}\left(p_{a}\right)$
decays faster than any power of $p_{a}$ as $p_{a}\rightarrow0$,
the integral $\int dp_{a}\frac{\left|\psi_{a\mid b}\left(p_{a}\right)\right|}{\left|p_{a}\right|}$
is finite. More concretely, $\left|\psi_{a\mid b}\left(p_{a}\right)\right|\sim\left|p_{a}\right|^{n}$
for all $n$ as $p_{a}\rightarrow0$. The finiteness of the gauge-invariant
expansion parameter is therefore already ensured by the request of having a
monotonic clock. As a result, the expansion scalar remains finite for
all $\tau$, which is a strong indicator of singularity resolution
and avoidance of the conditions of the Penrose-Hawking theorems. 

\subsection{Kretschmann scalar}\label{sec:Kretschmann}

The Kretschmann scalar $K=R_{\mu\nu\rho\sigma}R^{\mu\nu\rho\sigma}$
is one of the Riemann invariant whose divergence signals the presence
of similarities. It classical expression for our metric \eqref{eq:metric-ab}
becomes
\begin{equation}
\begin{aligned}K= & \frac{4}{a^{6}}\left(a+b\frac{p^{2}_{b}}{\alpha^{2}}\right)^{2}+\frac{2p^{2}_{b}}{a^{6}\alpha^{4}}\left(ap_{a}-bp_{b}\right)^{2}\\
 & +\frac{2}{N^{4}a^{6}}\left[2ab\left(\ddot{a}+\frac{\dot{N}}{\alpha}p_{b}\right)+\frac{N^{2}}{\alpha^{2}}p_{b}\left(ap_{a}-bp_{b}\right)\right]^{2}\\
 & +\frac{1}{N^{4}a^{6}}\left[a\left(a\ddot{b}-b\ddot{a}\right)+\frac{1}{\alpha}\left(ap_{a}-bp_{b}\right)\left(a\dot{N}-2\frac{N^{2}}{\alpha}p_{b}\right)\right]^{2}
\end{aligned}
\end{equation}
Using the definition of momenta $p_{a}=-\alpha\frac{\dot{b}}{N}$
and $p_{b}=-\alpha\frac{\dot{a}}{N}$ and the two EOM $\frac{d}{d\tau}\frac{\dot{a}}{N}=0=\frac{d}{d\tau}\frac{\dot{b}}{N}$
(see \cite{D_Ambrosio_2018}), the expression simplifies
\begin{equation}
K=\frac{4}{a^{6}}\left(a+b\frac{p^{2}_{b}}{\alpha^{2}}\right)^{2}+\frac{8p^{2}_{b}}{a^{6}\alpha^{4}}\left(ap_{a}-bp_{b}\right)^{2}.\label{eq:K-before-CH}
\end{equation}
The operator expression for the above
\begin{equation}
\hat{K}=4\widehat{\frac{1}{a^{4}}}+\frac{12}{\alpha^{4}}\hat{b}^{2}\hat{p}^{4}_{b}\widehat{\frac{1}{a^{6}}}+\frac{8}{\alpha^{2}}\hat{b}\hat{p}^{2}_{b}\widehat{\frac{1}{a^{5}}}-\frac{16}{\alpha^{4}}\hat{b}\hat{p}^{3}_{b}\hat{p}_{a}\widehat{\frac{1}{a^{5}}}+\frac{8}{\alpha^{4}}\hat{p}^{2}_{b}\hat{p}^{2}_{a}\widehat{\frac{1}{a^{4}}}
\end{equation}
is not symmetric and more importantly depends on both $\hat{b}$ and
$\hat{p}_{b}$ in addition to its dependence of the system variables
$\hat{a},\,\hat{p}_{a}$. Hence, once again we need to use Eqs. \eqref{eq:Re-expectation-general}
and \eqref{eq:expect-gauge-inv-obs-general-g} to derive the expectation
value of its gauge-invariant extension. Using these equations, and
considering $\hat{b}\left|p_{b}\right\rangle =i\hbar\frac{\partial}{\partial p_{b}}\left|p_{b}\right\rangle $,
and Eqs. \eqref{eq:Green-inverse-a} and \eqref{eq:Abs-int-condition},
we obtain 
\begin{equation}
\begin{aligned}\left|\left\langle \hat{F}_{K,T}\left(\tau\right)\right\rangle _{\text{phys}}\right|\leq & \frac{1}{\hbar^{4}}\int_{\mathbb{R}}dp^{\prime}_{a}\int_{\mathbb{R}}dp_{a}\,\left|\psi^{*}_{a|b}\left(p^{\prime}_{a}\right)\right|\left|\psi_{a|b}\left(p_{a}\right)\right|\left|\left(p^{\prime}_{a}-p_{a}\right)^{3}\right|\\
 & \times\Bigg|\frac{1}{3}+\frac{2}{3}\frac{p^{\prime2}_{a}}{p^{2}_{a}}\\
 & -\frac{\alpha^{4}}{20}\frac{\left(p^{\prime}_{a}-p_{a}\right)^{2}}{p^{4}_{a}}\left(\left(\frac{\partial g\left(p_{b}\right)}{\partial p_{b}}-\frac{\tau}{\hbar}\right)^{2}+i\frac{\partial^{2}g\left(p_{b}\right)}{\partial p^{2}_{b}}\right)_{p_{b}=-\frac{\alpha^{2}}{p_{a}}}\\
 & +i\frac{\alpha^{2}}{6}\frac{\left(p^{\prime}_{a}-p_{a}\right)}{p^{2}_{a}}\left(1+2\frac{p^{\prime}_{a}}{p_{a}}\right)\left(\frac{\partial g\left(p_{b}\right)}{\partial p_{b}}-\frac{\tau}{\hbar}\right)_{p_{b}=-\frac{\alpha^{2}}{p_{a}}}\Bigg|
\end{aligned}
\end{equation}
and using the triangle inequality we get
\begin{equation}
\begin{aligned}\left|\left\langle \hat{F}_{K,T}\left(\tau\right)\right\rangle _{\text{phys}}\right|\leq & \frac{1}{\hbar^{4}}\int_{\mathbb{R}}dp^{\prime}_{a}\int_{\mathbb{R}}dp_{a}\,\left|\psi^{*}_{a|b}\left(p^{\prime}_{a}\right)\right|\left|\psi_{a|b}\left(p_{a}\right)\right|\left|\left(p^{\prime}_{a}-p_{a}\right)^{3}\right|\\
 & \times\left\{ \frac{1}{3}+\frac{2}{3}\frac{p^{\prime2}_{a}}{p^{2}_{a}}\right.\\
 & +\frac{\alpha^{4}}{20}\frac{\left(p^{\prime}_{a}-p_{a}\right)^{2}}{p^{4}_{a}}\left(\left(\frac{\partial g\left(p_{b}\right)}{\partial p_{b}}-\frac{\tau}{\hbar}\right)^{2}+\left|\frac{\partial^{2}g\left(p_{b}\right)}{\partial p^{2}_{b}}\right|\right)_{p_{b}=-\frac{\alpha^{2}}{p_{a}}}\\
 & \left.+\frac{\alpha^{2}}{6}\frac{\left|p^{\prime}_{a}-p_{a}\right|}{p^{2}_{a}}\left|\left(1+2\frac{p^{\prime}_{a}}{p_{a}}\right)\right|\left|\left(\frac{\partial g\left(p_{b}\right)}{\partial p_{b}}-\frac{\tau}{\hbar}\right)_{p_{b}=-\frac{\alpha^{2}}{p_{a}}}\right|\right\} 
\end{aligned}
\end{equation}
Again we note that, since $\psi_{a\mid b}\in D_{0}\left(p_{a}\right)$
decays faster than any power of $p_{a}$ as $p_{a}\rightarrow0$ and
$p_a\to\pm\infty$, the integral above remains finite for all finite values of $\tau$.
Thus this equation establishes a strict upper bound on the physical
expectation value of the Kretschmann scalar. Consequently, for any
finite relational time $\tau$, the expectation value $\left|\left\langle \hat{F}_{K,T}(\tau)\right\rangle _{\text{phys }}\right|$
is bounded by a finite, purely quadratic polynomial in $\tau$. This
demonstrates that the Kretschmann scalar never diverges to infinity,
confirming that the classical curvature singularity is successfully
resolved in the quantum relational framework.

\section{Example: Gaussian states}\label{sec:gauss-states}

In this section we present some concrete results related to the general
framework we considered until now. To this end and to be able to perform
the computations analytically, we consider physical Gaussian states
in the following form
\begin{equation}
\psi^{(n)}_{a|b}\left(p_{a}\right)=N_{G}\Theta\left(-p_{a}\right)p^{n}_{a}e^{-\frac{p^{2}_{a}}{2\sigma^{2}}}\label{eq:gaussian-states-general}
\end{equation}
where $N_{G}$ is a normalization constant, $n\in\mathbb{N}\backslash\{0\}$, and
$\sigma$ is the width of the Gaussian. This quantum state is real,
$\psi^{(n)}_{a|b}=\psi^{(n)*}_{a|b}$ and represents the initial condition
of the black hole interior, in which the momentum $p_{a}$ is sharply
distributed around the only peak of the Gaussian at $\bar{p}_{a}=-\sigma\sqrt{n}<0$.
The fact that such states are centered around a negative value for
$p_{a}$ is consistent with the classical solutions. Recall that the
EOM resulting from Eq. \eqref{eq:action} yield
$p_{a}=\text{sgn}(\lambda)\alpha$, and because the black hole solution
corresponds to the case in which $\lambda\in[-\sqrt{2GM},0)$, we
conclude $p_{a}=-\alpha<0$ in the black hole region. Moreover, such
states have null support in $p_{a}=0$, which was one of the fundamental
requirements for the construction of the physical Hilbert space, which
also assures the monotonicity of the quantum clock.

The states \eqref{eq:gaussian-states-general} are not the exact wave
functions of \eqref{eq:Thiemann-funct} belonging to $D_{0}\left(p_{a}\right)$
in \eqref{eq:D0-pa}. However, they belong to the closure of their
span, i.e, the same physical Hilbert space $\mathscr{H}_{\text{ph}ys}=\overline{D_{0}\left(p_{a}\right)}=L^{2}\left(\mathbb{R},dp_{a}\right)$.
Moreover, they share the relevant features such as rapid decay for
$p_{a}\to0$ and $p_{a}\to\pm\infty$, ensuring the convergence of
the integrals involved with a suitable choice of $n \in \mathbb{N}\backslash\{0\}$. Also, notice that the states \eqref{eq:gaussian-states-general}
allow the case for which \textbf{$a=0$}. This is seen by writing
the Fourier transform of $\psi^{(n)}_{a|b}\left(p_{a}\right)$ as
\begin{equation}
\tilde{\psi}^{(n)}_{a|b}\left(a\right)=\int_{\mathbb{R}}dp_{a}N_{G}\Theta\left(-p_{a}\right)p^{n}_{a}e^{-\frac{p^{2}_{a}}{2\sigma^{2}}}e^{\frac{i}{\hbar}ap_{a}}
\end{equation}
which implies
\begin{equation}
\begin{aligned}\tilde{\psi}^{(n)}_{a|b}\left(0\right)= & \int_{\mathbb{R}}dp_{a}\,N_{G}\Theta\left(-p_{a}\right)p^{n}_{a}e^{-\frac{p^{2}_{a}}{2\sigma^{2}}}\\
= & N_{G}(-1)^{n}2^{\frac{n-1}{2}}\sigma^{n+1}\Gamma\left(\frac{n+1}{2}\right)\\
\neq & 0
\end{aligned}
\label{eq:a0-yes}
\end{equation}
Thus the results that we will obtain in this section hold even if
the black hole is in a quantum state with $a=0$, which classically
corresponds to a singularity.

Since 
\begin{equation}
\psi_{\text{kin}}\left(p_{a},p_{b}\right)\bigg|_{C_{H}=0}=\psi_{\text{kin}}\left(p_{a},-\frac{\alpha^{2}}{p_{a}}\right)=\psi_{a|b}\left(p_{a}\right)
\end{equation}
we can express such a kinematical state as
\begin{equation}
\psi^{(n)}_{\text{kin}}\left(p_{a},p_{b}\right)=N_{G}\Theta\left(-p_{a}\right)p^{n}_{a}e^{-\frac{p^{2}_{a}}{2\sigma^{2}}}e^{-\frac{\left(p_{b}+\frac{\alpha^{2}}{p_{a}}\right)^{2}}{2\delta^{2}}}\in L^{2}(\mathbb{R}^{2},dp_{a}dp_{b})=\mathscr{H}_{\text{kin}},
\end{equation}
for some parameter $\delta\in\mathbb{R}$. We can also see that 
\begin{equation}
\psi^{(n)}_{\text{kin}}\left(p_{a},p_{b}\right)\bigg|_{C_{H}=0}=\psi^{(n)}_{\text{kin}}\left(p_{a},-\frac{\alpha^{2}}{p_{a}}\right)=N_{G}\Theta\left(-p_{a}\right)p^{n}_{a}e^{-\frac{p^{2}_{a}}{2\sigma^{2}}}=\psi^{(n)}_{a|b}\left(p_{a}\right)
\end{equation}
as desired.

\subsection{Area of 2-spheres }\label{subsec:Gaussian-Area}

To study the area of 2-spheres using the states \eqref{eq:gaussian-states-general},
we choose the quantum relational clock operator to be $\hat{T}=\hat{b}$,
or equivalently, $g(p_{b})=0$. Using this condition, and looking
at the first two lines of \eqref{eq:F-Ab-phys-phys-explicit} and
recalling that the states are real, $\psi^{(n)*}_{a|b}=\psi^{(n)}_{a|b}$,
we obtain
\begin{equation}
\begin{aligned}\left\langle \hat{F}_{A,T}\left(\tau\right)\right\rangle _{\text{phys}}= & 4\pi\int_{\mathbb{R}}dp_{a}\,\frac{\alpha^{4}}{p^{4}_{a}}\tau^{2}\left|\psi^{(n)}_{a|b}\left(p_{a}\right)\right|^{2}+4\pi\hbar^{2}\int_{\mathbb{R}}dp_{a}\,\left|\frac{\partial\psi^{(n)}_{a|b}\left(p_{a}\right)}{\partial p_{a}}\right|^{2}\end{aligned}
.\label{eq:GIQArea}
\end{equation}
Given the form of the first term in the above, the states \eqref{eq:gaussian-states-general}
that will yield well-defined results are the ones with $n\geq2$,
so we will restrict our attention to this case throughout this section.

 It is evident from Eq. \eqref{eq:GIQArea} that at $\tau=0$ one obtains
the minimum area of 2-spheres
\begin{equation}
A_{\text{min}}=4\pi\hbar^{2}\int_{\mathbb{R}}dp_{a}\,\left|\frac{\partial\psi^{(n)}_{a|b}\left(p_{a}\right)}{\partial p_{a}}\right|^{2}\label{eq:min-area-Gaussian}
\end{equation}
which is strictly positive, even if the state corresponds to $a=0$,
as discussed in Eq. \eqref{eq:a0-yes}. As we will see below, the
nonvanishing of this minimum area is a direct consequence of the quantum
fluctuations resulted from the uncertainty in $\hat{a}$. This can
be seen more clearly by writing the minimum area \eqref{eq:min-area-Gaussian}
in terms of the uncertainty in $\hat{a}$, namely, $\Delta a=\sqrt{\left\langle \hat{a}^{2}\right\rangle -\left\langle \hat{a}\right\rangle ^{2}}$.
In the momentum representation $\hat{a}=i\hbar\,\frac{\partial}{\partial p_{a}}$,
we ca wite
\begin{equation}
\left\langle \hat{a}^{2}\right\rangle =\int_{\mathbb{R}}dp_{a}\,\psi^{(n)*}_{a|b}\left(p_{a}\right)\left(-\hbar^{2}\frac{\partial^{2}}{\partial p^{2}_{a}}\right)\psi^{(n)}_{a|b}\left(p_{a}\right)=\hbar^{2}\int_{\mathbb{R}}dp_{a}\,\left|\frac{\partial\psi^{(n)}_{a|b}\left(p_{a}\right)}{\partial p_{a}}\right|^{2},
\end{equation}
where boundary terms have been assumed to vanish due to decay of $\psi^{(n)}_{a|b}(p_{a})$
at infinities. On other hand, since the states \eqref{eq:gaussian-states-general}
are real, we also have the well-known result
\begin{equation}
\left\langle \hat{a}\right\rangle =\int_{\mathbb{R}}dp_{a}\,\psi^{(n)*}_{a|b}\left(p_{a}\right)i\hbar\frac{\partial\psi^{(n)}_{a|b}\left(p_{a}\right)}{\partial p_{a}}=\frac{i\hbar}{2}\int_{\mathbb{R}}dp_{a}\frac{\partial}{\partial p_{a}}\left(\psi^{(n)}_{a|b}\left(p_{a}\right)\right)^{2}=0
\end{equation}
where we have again assumed that the states vanish at the boundaries.
From these it is evident that we can write 
\begin{equation}
A_{\text{min}}=4\pi\left(\Delta a\right)^{2}.\label{eq:area-Delta-a}
\end{equation}
This very illuminating result shows that the minimum non-zero area
is rooted in quantum effects as expected. This is also consistent
with what we showed in Eq. \eqref{eq:psi-ab-zero-area}, that it is
not possible to have $\Delta a=0$, as it will result in non-normalizable
states, i.e., states that do not belong to the physical Hilbert space
$\mathscr{H}_{\text{phys}}$.

To find a more explicit expression for \eqref{eq:min-area-Gaussian},
we first find the normalization constant $N_{G}$ using the normalization
condition $\int_{\mathbb{R}}dp_{a}|\psi^{(n)}_{a|b}(p_{a})|^{2}=1$,
as 
\begin{equation}
N_{G}=\sqrt{\frac{2}{\sigma^{2n+1}\Gamma\left(n+\frac{1}{2}\right)}}
\end{equation}
where we have used
\begin{equation}
\int^{\infty}_{0}dp_{a}\,p^{2k}_{a}e^{-\frac{p^{2}_{a}}{\sigma^{2}}}=\frac{\sigma^{2k+1}}{2}\Gamma\left(k+\frac{1}{2}\right).\label{Eq.Semi-Integral-NG}
\end{equation}
Replacing this back into \eqref{eq:gaussian-states-general} and plugging
it in \eqref{eq:min-area-Gaussian} yields 
\begin{equation}
\begin{split}A_{\text{min}}=4\pi\hbar^{2}\frac{1}{\sigma^{2}}\left(\frac{4n-1}{4n-2}\right).\end{split}
\label{eq:area-min-gaussian-explicit}
\end{equation}
Once again we can express the above result in terms of the uncertainty
in configuration variable $a$. Using $\hat{a}=i\hbar\frac{\partial}{\partial p_{a}}$
and computing the corresponding uncertainty over the states \eqref{eq:gaussian-states-general}
we obtain
\begin{equation}
\begin{aligned}\Delta a= & \sqrt{\left\langle \hat{a}^{2}\right\rangle -\left\langle \hat{a}\right\rangle ^{2}}\end{aligned}
=\frac{\hbar}{\sigma}\sqrt{\frac{4n-1}{4n-2}}.
\end{equation}
which again leads to \eqref{eq:area-Delta-a}.

In Appendix \ref{sec:app-general-clock-min-area}, we will show
that the result $A_{\text{min}}>0$ for Gaussian states holds not
only for the specific clock choice $\hat{T}=\hat{b}$, but for all
$\hat{T}$ that satisfy $[\hat{T},\hat{H}_{C}]=i\hbar$ which are
defined up to $\hat{T}\rightarrow\hat{T}+h(\hat{H}_{C})$. We will
also extend such result to any real state $\psi_{a|b}(p_{a})$, without any
particular restriction to Gaussian functions.

Another observation about $A_{\text{min}}$ is the following. We stated
before that the peak of the Gaussian \eqref{eq:gaussian-states-general}
is at $\bar{p}_{a}\propto\sigma$ and thus $\sigma$ has dimensions
of momentum. Using 
\begin{equation}
l_{P}=\sqrt{\hbar G},\quad p_{P}=\frac{\hbar}{l_{P}}=\sqrt{\frac{\hbar}{G}}
\end{equation}
in units where speed of light $c=1$ and replacing one $\hbar$
from $\ell_{p}$ and the other $\hbar$ from $p_{p}$ in \eqref{eq:area-min-gaussian-explicit}
yields
\begin{equation}
A_{\min}=4\pi\ell^{2}_{p}\left(\frac{p_{p}}{\sigma}\right)^{2}\left(\frac{4n-1}{4n-2}\right),
\end{equation}
where $4\pi\ell^{2}_{p}=A_{p}$ is the Planck area. From this we see that, for a spread $\sigma\gg p_{p}$, the minimum
area goes to zero. In the same way, for large $n$
and $\sigma\sim p_{p}$, we get $A_{\min}=A_{p}$.

If one wants to set the area \eqref{eq:area-min-gaussian-explicit}
equal to the loop quantum gravity area gap, $A_{\text{LQG}}=8\pi\gamma l^{2}_{P}$,
one obtains
\begin{equation}
\sigma=\frac{\hbar}{l_{P}}\sqrt{\frac{1}{2\gamma}\left(\frac{n-\frac{1}{4}}{n-\frac{1}{2}}\right)},
\end{equation}
which for large $n$ reduces to 
\begin{equation}
\sigma=\frac{\hbar}{\sqrt{2\gamma}l_{P}}=\frac{p_P}{\sqrt{2\gamma}}.
\end{equation}

\subsection{Expansion scalar}

Let us turn our attention to the expansion scalar \eqref{eq:F-theta-phys-exp-fin}.
Choosing $g=0$ or equivalently $\hat{T}=\hat{b}$, we obtain 
\begin{equation}
\begin{aligned}\mathfrak{Re}\left\langle \hat{F}_{\vartheta,T}\left(\tau\right)\right\rangle _{\text{phys}}= & \mathfrak{Re}\left\{ \frac{i\alpha}{\hbar}N^{2}_{G}\int^{\infty}_{0}dp^{\prime}_{a}\int^{\infty}_{0}dp_{a}\,\left(-1\right)^{2n-1}p^{\prime n}_{a}p^{n-1}_{a}\right.\\
 & \left.\times e^{-\frac{1}{2\sigma^{2}}\left(p^{2}_{a}+p^{\prime2}_{a}\right)}e^{-\frac{i}{\hbar}\left(-\frac{\alpha^{2}}{p_{a}}+\frac{\alpha^{2}}{p^{\prime}_{a}}\right)\tau}\text{sgn}\left(p^{\prime}_{a}-p_{a}\right)\right\} 
\end{aligned}
\label{eq:theta-Gaussian}
\end{equation}
Using Euler's formula $e^{i\phi}=\cos\left(\phi\right)+i\sin\left(\phi\right)$
in the above yields
\begin{equation}
\begin{aligned}\mathfrak{Re}\left\langle \hat{F}_{\vartheta,T}\left(\tau\right)\right\rangle _{\text{phys}}= & \frac{\alpha}{\hbar}N^{2}_{G}\int^{\infty}_{0}dp^{\prime}_{a}\int^{\infty}_{0}dp_{a}\,\left(-1\right)^{2n-1}p^{\prime n}_{a}p^{n-1}_{a}\\
 & \times e^{-\frac{1}{2\sigma^{2}}\left(p^{2}_{a}+p^{\prime2}_{a}\right)}\left[\sin\left(\frac{\alpha^{2}\tau}{\hbar}\left(\frac{1}{p^{\prime}_{a}}-\frac{1}{p_{a}}\right)\right)\right]\\
 & \times\text{sgn}\left(p^{\prime}_{a}-p_{a}\right)
\end{aligned}
\label{eq:theta-sin-odd}
\end{equation}
It is immediately seen from the above that 
\begin{equation}
\mathfrak{Re}\left\langle \hat{F}_{\vartheta,T}\left(\tau=0\right)\right\rangle _{\text{phys}}=0.\label{eq:theta-at0-0}
\end{equation}
Moreover, the derivative of the expansion scalar is
\begin{equation}
\begin{aligned}\frac{d}{d\tau}\mathfrak{Re}\left\langle \hat{F}_{\vartheta,T}\left(\tau\right)\right\rangle _{\text{phys}}= & \frac{\alpha}{\hbar}N^{2}_{G}\int^{\infty}_{0}dp^{\prime}_{a}\int^{\infty}_{0}dp_{a}\,\left(-1\right)^{2n-1}p^{\prime n}_{a}p^{n-1}_{a}\\
 & \times e^{-\frac{1}{2\sigma^{2}}\left(p^{2}_{a}+p^{\prime2}_{a}\right)}\left[\cos\left(\frac{\alpha^{2}\tau}{\hbar}\left(\frac{1}{p^{\prime}_{a}}-\frac{1}{p_{a}}\right)\right)\right]\\
 & \times\frac{\alpha^{2}}{\hbar}\left(\frac{1}{p^{\prime}_{a}}-\frac{1}{p_{a}}\right)\text{sgn}\left(p^{\prime}_{a}-p_{a}\right).
\end{aligned}
\end{equation}
Given the limits of the integral, the term $p^{\prime n}_{a}p^{n-1}_{a}$
in the integrand above is strictly positive, while the term $\left(-1\right)^{2n-1}$
is strictly negative. The last line can be written as $-\frac{\left|p^{\prime}_{a}-p_{a}\right|}{p_{a}p^{\prime}_{a}}$
which again given the integral limits is strictly negative. Hence,
at $\tau=0$ we get
\begin{equation}
\frac{d}{d\tau}\mathfrak{Re}\left\langle \hat{F}_{\vartheta,T}\left(\tau\right)\right\rangle _{\text{phys}}>0.\label{eq:theta-dot-at0-0}
\end{equation}
The two results \eqref{eq:theta-at0-0} and \eqref{eq:theta-dot-at0-0}
show that at $\tau=0$, which corresponds to the classical singularity,
the expectation value of the physical expansion scalar vanishes and
thus the singularity does not exist anymore. Moreover, these two equations
show that a bounce from the black hole spacetime to the white hole
spacetime happens at $\tau=0$. Furthermore, we see from \eqref{eq:theta-sin-odd}
that since $\sin$ is an odd function of $\tau$ we can conclude
\begin{equation}
\mathfrak{Re}\left\langle \hat{F}_{\vartheta,T}\left(\tau\right)\right\rangle _{\text{phys}}=-\mathfrak{Re}\left\langle \hat{F}_{\vartheta,T}\left(-\tau\right)\right\rangle _{\text{phys}}.
\end{equation}
This result further affirms that $\tau=0$ corresponds to the bounce
from the a trapped to an antitrapped region.

\begin{figure}
\centering \includegraphics[scale=0.65]{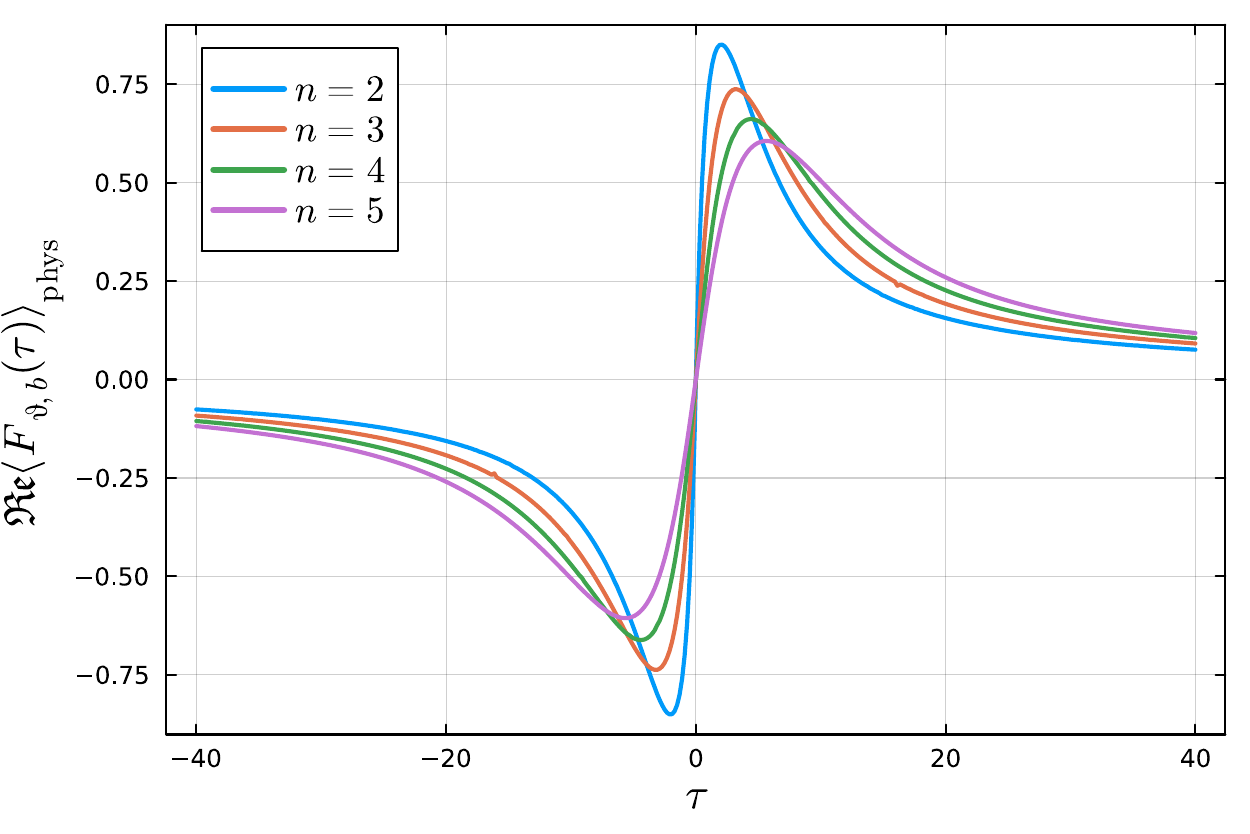}
\caption{Expectation value of the (gauge-invariant) expansion parameter evolution,
for the quantum clock $\hat{T}=\hat{b}$ and the state $\psi^{(n)}_{a|b}\left(p_{a}\right)=N_{G}\Theta\left(-p_{a}\right)p^{n}_{a}e^{-\frac{p^{2}_{a}}{2\sigma^{2}}}$.
In the specific case presented here, for the plot we have used $\hbar=\sigma=\alpha=1$.}
\label{exp_param_plot_multiple}
\end{figure}

It is also illuminating to study the asymptotic behavior of the gauge-invariant
expansion scalar for $\tau\to\pm\infty$. For this, we will employ
the Riemann-Lebesgue lemma which states that, given an integrable function
$f\in L^{1}[0,\infty)$ such that $\int_{\mathbb{R}^{n}}|f(x)|dx<\infty$,
its Fourier transform vanishes at infinity, i.e., $\left|\int_{\mathbb{R}^{n}}f(x)\mathrm{e}^{-\mathrm{i}x\xi}dx\right|\to0$
as $\left|\xi\right|\to\infty$. Let us look at the $p_{a}$ integral
in \eqref{eq:theta-Gaussian}
\begin{equation}
\begin{aligned}I_{p_{a}}= & \int^{\infty}_{0}dp_{a}\,p^{n-1}_{a}e^{-\frac{p^{2}_{a}}{2\sigma^{2}}}e^{i\frac{\alpha^{2}}{\hbar p_{a}}\tau}\text{sgn}\left(p^{\prime}_{a}-p_{a}\right)\end{aligned}
\label{eq:I-pa}
\end{equation}
Making a substitution $x=\frac{\alpha^{2}}{\hbar p_{a}}\Rightarrow dx=-\frac{\alpha^{2}}{\hbar p^{2}_{a}}dp_{a}$
yields
\begin{equation}
\begin{aligned}I_{p_{a}}= & \int^{\infty}_{0}dx\,\left(\frac{\alpha^{2}}{\hbar}\right)^{n}\frac{1}{x^{n+1}}e^{-\frac{\alpha^{4}}{2\hbar^{2}x^{2}\sigma^{2}}}\text{sgn}\left(p^{\prime}_{a}-\frac{\alpha^{2}}{\hbar x}\right)e^{ix\tau}\end{aligned}
.
\end{equation}
The function $\left(\frac{\alpha^{2}}{\hbar}\right)^{n}\frac{1}{x^{n+1}}e^{-\frac{\alpha^{4}}{2\hbar^{2}x^{2}\sigma^{2}}}\text{sgn}\left(p^{\prime}_{a}-\frac{\alpha^{2}}{\hbar x}\right)$
belongs to $L^{1}[0,\infty)$ for each of its branches corresponding
to $\text{sgn}\left(p^{\prime}_{a}-\frac{\alpha^{2}}{\hbar x}\right)$
being $\pm1$, i.e., 
\begin{equation}
\left|\int^{\infty}_{0}dx\,\left(\frac{\alpha^{2}}{\hbar}\right)^{n}\frac{1}{x^{n+1}}e^{-\frac{\alpha^{4}}{2\hbar^{2}x^{2}\sigma^{2}}}\right|<\infty.
\end{equation}
This is because at $x\to0^{+}$ the Gaussian $e^{-\frac{\alpha^{4}}{2\hbar^{2}x^{2}\sigma^{2}}}\to0$
faster than any power of $x$ so the integral vanishes, and at $x\to\infty$
the term $x^{-(n+1)}\rightarrow0$ for $n\geq0$. Hence, by Riemann-Lebesgue
lemma we conclude that its Fourier transform \eqref{eq:I-pa} decays
to zero as $\tau\to\pm\infty$, which consequently means that 
\begin{equation}
\lim_{\tau\to\pm\infty}\mathfrak{Re}\left\langle \hat{F}_{\vartheta,T}\left(\tau\right)\right\rangle _{\text{phys}}=0.
\end{equation}
So far we have obtained the behavior of the expansion scalar at $\tau\to0$
and $\tau\to\pm\infty$, and have noticed that it is an odd function
in $\tau$. Computing an explicit expression for the
expansion parameter for a general $\tau$, however, is not possible
given the complicated form of the integral \eqref{eq:theta-Gaussian}.
One can, however, compute this integral numerically. The plot of the
numerical computation of the expansion scalar is presented in Fig.
\ref{exp_param_plot_multiple}. It is clearly seen from this plot
that the expansion parameter inside the black hole for finite negative
values of $\tau$ is negative, but near the deep quantum regime, i.e.,
$\tau\to0$, it reverses behavior and vanishes at $\tau=0$. This
is distinctly different from the classical behavior where the expansion
scalar goes to $-\infty$ as $\tau\to0$, where the singularity resides.
Moreover, in the quantum theory, the expansion scalar bounces to positive
values and remains positive and finite for finite positive values
of $\tau$, which is the characteristic behavior of an anti-trapped
region or a white hole.


It is worth mentioning that the same behavior as the one in Fig. \ref{exp_param_plot_multiple}
can be reproduced as $\vartheta\left(\tau\right)$ by considering
an area function $A=A_{\text{min}}+\beta\tau^{2}$ whose expansion
becomes
\begin{equation}
\vartheta\left(\tau\right)=\frac{1}{A\left(\tau\right)}\frac{dA\left(\tau\right)}{d\tau}=\frac{2\beta\tau}{A_{\text{min}}+\beta\tau^{2}},
\end{equation}
where $A_{\text{min}}>0$ and $\beta\in\mathbb{R}$ is some acceleration
parameter. This also shows that the classical behavior (when $A_{\text{min}}=0$
for $\hbar\to0$; see Eq. \eqref{eq:area-min-gaussian-explicit})
yields $\vartheta\left(\tau\right)\propto\frac{1}{\tau}$ which is
the qualitative behavior one expects from a classical expansion scalar
in the interior of the Schwarzschild black hole \cite{Fragomeno:2024tlh, Gingrich:2024mgk}. 

These results explicitly show that the singularity is resolved in
this framework as a direct consequence of quantum effects, including
the uncertainty in metric components, i.e., in $a$, and associated
quantum fluctuations.

\section{Conclusions\label{sec:Conclusion}}

In this work, we have studied the dynamics of the interior of a Schwarzschild
black hole, which is isometric to the Kantowski-Sachs metric, in the
quantum relational formalism. The model has a four dimensional phase
space with configuration variables $a,\,b$ and conjugate momenta
$p_{a},\,p_{b}$. Since the Hamiltonian of the model can be written
in a separable way as the Hamiltonian of the clock plus the Hamiltonian
of the system, i.e., $C_{H}=p_{b}+\frac{\alpha^{2}}{p_{c}}$, we were
able to implement the Page-Wootters formalism and construct quantum
clocks corresponding to $b$. The physical Hilbert space of the model
$\mathscr{H}_{\mathrm{phys}}\cong L^{2}\left(\mathbb{R},dp_{a}\right)$
was then constructed using the refined algebraic quantization. The
physical states were written down using Neuser-Thiemann's states \cite{thiemann2023propertiessmoothdenseinvariant,neuser2023smoothinvariantorthonormalbasis} which decay fast enough at $p_{a}\to0$ and $p_{a}\to\pm\infty$. This ensured
the monotonicity of the quantum clock and allowed the Hamiltonian
constraint to have a well-defined action in the standard Schrodinger
representation.

After setting this stage, we studied the general form and properties
of the expectation values of three gauge-invariant quantum observables,
pertinent to the singularity resolution, namely the area of 2-speheres,
the expansion scalar, and the Kretschmann scalar. We computed these
expectation values as a function of the relational time $\tau$.

For the expectation area of 2-spheres inside the black hole, we showed
that i) it follows a quadratic evolution equation in $\tau$ which
points to a black-to-white hole bounce, ii) it has a nonzero minimum
$A_{\text{min}}$. 

For the expansion scalar of radial null geodesics, we found results
consistent with the above. In particular, we found that it is finite
everywhere in the interior.

Similarly, we found that the bound on the expectation value of the Kretschmann
scalar is also quadratic in $\tau$, and thus the Kretschmann is finite everywhere inside
the black hole. 

These general results imply that the singularity of the black hole
is resolved in this approach. To show these results in a more concrete
way, we chose specific physical states which were real and Gaussian,
and computed the above expectation values specific for such states.

Our analysis of the area of 2-spheres for these states revealed that
the minimum area is actually proportional to the uncertainty in $a$,
namely the quantum fluctuations, and the associated uncertainty in
$a$ prohibits the minimum area to vanish.

We also studied the expansion scalar for these states and analytically showed that
it vanishes at $\tau\to\pm\infty$ and at $\tau=0$, where its derivative
with respect to $\tau$ is positive. Moreover, it is an odd function of
$\tau$ around $\tau=0$ and never diverges inside the black hole.
These results were also confirmed by directly computing the expansion
scalar numerically. Thus, once again we confirmed that the singularity
of the black hole is resolved, and there is a bounce from the black
hole to a white hole at $\tau=0$, which used to be a classical singularity.

As a result of all of the above observations, we conclude that applying
the relational framework and the usual Schrodinger representation
to the interior of the Schwarzschild black hole resolves its singularity.


\acknowledgments{
We are grateful to Kristina Giesel for valuable discussions on physical Hilbert space of our model. We also thank Alex Koek for his help in finding the classical expression for the Kretschmann scalar, and Carlo Rovelli for discussions and his insight about the metric and relational dynamics. S. R. acknowledges the support of the Natural Science and Engineering Research Council of Canada, funding reference No. RGPIN-2021-03644 and No. DGECR-2021-00302.
}







\appendix 

\section{Review of RAQ\label{app:RAQ-Brief}}

In constrained systems with continuous spectrum, a set of first class
constraints $\hat{C}_{I}$ and a kinematical unconstrained square
integrable Hilbert space $\mathscr{H}_{\text{kin}}$, the physical
states would be the ones that satisfy $\hat{C}_{I}\left|\psi\right\rangle =0$.
As mentioned above, for such systems with $0$ eigenvalue, the physical
states lie outside of $\mathscr{H}_{\text{kin}}$. An example is a
system whose momentum $p$ is constrained to vanish. Therefore, physical
states satisfy $\hat{p}\Psi_{\text{phys}}(p)=p\Psi_{\text{phys}}(p)=0$
and thus we get $\Psi_{\text{phys}}(p)=\delta(p)$. This is a distribution
and is not normalizable, so it lies outside of $\mathscr{H}_{\text{kin}}$.
To find the physical states which lie in a larger space, one applies
RAQ.

The arena of RAQ includes three spaces, the so called, Gelfand triple
\begin{equation}
\Phi\subset\mathscr{H}_{\text{kin}}\subset\Phi^{*},\label{eq:Gelfand-general}
\end{equation}
where
\begin{itemize}
\item $\mathscr{H}_{\text{kin}}$ is the kinematical Hilbert space of square
integrable functions on which the classical unconstrained system is
represented, 
\item $\Phi\subset\mathscr{H}_{\text{kin}}$ is a dense subspace of test
states whose states are infinitely differentiable and all derivatives
rapidly decay (nuclear Fr\'{e}chet space), which is a ``seed''
space for physical states,
\item $\Phi^{*}$ is the topological dual of $\Phi$, consisting of all
continuous linear functionals on $\Phi$. Physical states live in
$\Phi^{*}$.
\end{itemize}
RAQ has the following steps:
\begin{enumerate}
\item \label{enu:RAQ-step-1}Kinematical Hilbert space: 
\begin{enumerate}
\item \textbf{States:} This is a square integrable space 
\begin{equation}
L^{2}(X,d\mu)=\left\{ \psi:X\rightarrow\mathbb{C}\vert\:\int_{X}\psi^{*}\left(x\right)\psi\left(x\right)d\mu\left(x\right)<\infty\right\} .
\end{equation}
\item \textbf{Operators:} The classical algebra of the unconstrained system
is represented on the kinematical Hilbert space $\mathscr{H}_{\text{kin}}$.
Classical operators and constraints $C_{I}$ are represented as self-adjoint
operators $\hat{C}_{I}$ on $\mathscr{H}_{\text{kin}}$. 
\item \textbf{Inner product:} The kinematical inner product is defined as
\begin{equation}
\left\langle \psi_{1}|\psi_{2}\right\rangle _{\text{kin}}=\int_{X}\psi^{*}_{1}(x)\psi_{2}(x)d\mu(x).
\end{equation}
\end{enumerate}
\item \label{enu:RAQ-step-2}Space of test functions $\Phi$: 
\begin{enumerate}
\item \textbf{States:} elements (or test states) of a dense subspace, $\Phi\subset\mathscr{H}_{\text{kin}}$,
which is a nuclear Fr\'{e}chet space. The functions are typically
infinitely differentiable and fall off quickly at infinity and possibly
other points, e.g., at zero. 
\item \textbf{Operators:} If $\hat{O}$ is an operator densely defined on
$\mathscr{H}_{\text{kin }}$, the corresponding operator acting on
$\Phi$ is then denoted by $\hat{O}^{\prime}$ and its action on $\Phi$
is defined by restricting its domain to $\Phi$.  We also require
$\Phi$ to have three properties: be invariant under the action of
the quantum constraint algebra, i.e., $\hat{C}_{I}\phi\in\Phi$ for
all $\phi\in\Phi$, be stable under the observable algebra $\hat{O}\phi\in\Phi,\,\forall\phi\in\Phi$
for $\left[\hat{O},\hat{C}_{I}\right]=0$ on $\Phi$, and be equipped
with a finer topology than $\mathscr{H}_{\text{kin}}$. The action
of any operator $\hat{O}$ should not take us out of $\Phi$%
, introduce singularities, or ruin the decay properties of the state. 
\item \textbf{Inner product:} operationally we use the same inner product
on $\Phi$, as the one on $\mathscr{H}_{\text{kin }}$.
\end{enumerate}
\item \label{enu:RAQ-step-3}Space of linear functionals $\Phi^{*}$:
\begin{enumerate}
\item \textbf{States:} The space $\Phi^{*}$ is the space of all continuous
linear functionals, $\Psi:\Phi\to\mathbb{C}$, on $\Phi$. For a state
$\phi\in\Phi$, a corresponding $\Psi_{\phi}\in\Phi^{*}$ is a linear
functional 
\begin{align}
\Psi_{\phi}: & \Phi\to\mathbb{C}\nonumber \\
 & f\mapsto\langle\phi\mid f\rangle_{\text{kin }}
\end{align}
in $\Phi^{*}$ where 
\begin{equation}
\Psi_{\phi}(f):=\langle\phi\mid f\rangle_{\text{kin }}=\int_{X}\phi^{*}\left(x\right)f\left(x\right)\,d\mu\left(x\right).
\end{equation}
Note that not all elements in $\Phi^{*}$ look like $\Psi_{\phi}$
since $\Phi^{*}$ is strictly larger than $\Phi$ (and $\mathscr{H}_{\text{kin }}$)
and contains distributions, such as the Dirac delta, that cannot be
written as $\langle\phi\mid\cdot\rangle_{\text{kin }}$ for any square-integrable
$\phi$.
\item \textbf{Operators:} The action of $\hat{O}$ or equivalently $\hat{O}^{\prime}$
on $\Phi^{*}$ is defined, under certain conditions\footnote{One defines a $\star$ that mimics the Hermitian conjugate operation
for observables, and then $\left(\hat{A}^{\prime}\Psi\right)[\phi]=\Psi\left[\hat{A}^{\star}\phi\right]$.
But if the representation $\pi$ of operators on $\Phi$ is faithful,
i.e., $\pi\left(A^{\star}\right)=\pi(A)^{\dagger}$, then $\left(\hat{A}^{\prime}\Psi\right)[\phi]=\Psi\left[\hat{A}^{\dagger}\phi\right]$}, as 
\begin{equation}
\left(\hat{O}\Psi_{\phi}\right)[f]\coloneqq\Psi\left[\hat{O}^{\prime\dagger}f\right],\,\forall\phi,f\in\Phi,\Psi\in\Phi^{*}.
\end{equation}
\item \textbf{Inner product:} There is no global inner product on the entirety
of $\Phi^{*}$ and in fact we do not need one (see next item).
\end{enumerate}
\item \label{enu:RAQ-step-4}Physical Hilbert space $\mathscr{H}_{\text{phys}}$: 
\begin{enumerate}
\item \textbf{States:} The physical states are defined as distributions
$\Psi_{\text{phys}}\in\Phi^{*}$, such that assuming self-adjointness,
$\hat{C}^{\prime\dagger}_{I}=\hat{C}^{\prime}_{I}$,
\begin{equation}
\left(\hat{C}_{I}\Psi^{\text{phys}}_{\phi}\right)\left[f\right]=\Psi^{\text{phys}}_{\phi}\left[\hat{C}^{\prime\dagger}_{I}f\right]=\Psi^{\text{phys}}_{\phi}\left[\hat{C}^{\prime}_{I}f\right]=0,\quad\forall\phi,f\in\Phi.
\end{equation}
Note that for non-self-adjoint operators, for this to be mathematically
well-defined, the kinematic adjoint $\hat{O}^{\prime\dagger}$ must
leave the test space invariant, $\hat{O}^{\dagger}\Phi\subset\Phi$.
In any case, we denote the space of all physical solutions as 
\begin{equation}
\mathcal{V}_{\text{phys }}\coloneqq\left\{ \Psi^{\text{phys}}_{\phi}\in\Phi^{*}\mid\left(\hat{C}_{I}\Psi_{\text{phys}}\right)[f]=0,\,\forall\phi,f\in\Phi,\forall I\right\} .
\end{equation}
These physical states are defined as
\begin{equation}
\begin{aligned}\Psi^{\text{phys}}_{\phi}: & \Phi\to\mathbb{C}\\
 & f\mapsto\eta(\phi)[f]
\end{aligned}
\end{equation}
where, given $\hat{U}(g)=e^{i\lambda^{I}\hat{C}_{I}}$, the unitary
representation of the group $G$ of the constraint Lie algebra with
$\lambda^{I},\,I=1,\ldots,N$ being real-valued group coordinates/parameters,
we have
\begin{equation}
\Psi^{\text{phys}}_{\phi}\left[f\right]\coloneqq\eta(\phi)[f]=\frac{1}{\left(2\pi\right)^{N}}\int_{G}d\mu(g)\langle\hat{U}(g)\phi\mid f\rangle_{\mathrm{kin}}.\label{eq:rigging-map}
\end{equation}
The above map 
\begin{equation}
\begin{aligned}\eta: & \Phi\rightarrow\mathcal{V}_{\text{phys }}\\
 & \phi\mapsto\Psi^{\text{phys}}_{\phi}\left[\cdot\right]
\end{aligned}
\end{equation}
is called a rigging map. It takes a kinematic test state and projects
it onto a physical solution. One can show that unimodularity of $G$
implies that $\eta$ is compatible with the adjointness relations
of the observable algebra, and $\left\langle \Psi^{\text{phys}}_{\phi_{1}}\mid\Psi^{\text{phys}}_{\phi_{2}}\right\rangle _{\text{phys}}=\left\langle \Psi^{\text{phys}}_{\phi_{2}}\mid\Psi^{\text{phys}}_{\phi_{1}}\right\rangle ^{*}_{\text{phys}}$
or equivalently $\eta\left(\phi_{1}\right)\left[\phi_{2}\right]=\eta\left(\phi_{2}\right)\left[\phi_{1}\right]^{*}$.
Note that $\eta(\phi)[\cdot]$ defined as
\begin{equation}
\begin{aligned}\eta(\phi): & \Phi\to\mathbb{C}\\
 & f\mapsto\Psi^{\text{phys}}_{\phi}\left[f\right]
\end{aligned}
\end{equation}
is an element of $\mathcal{V}_{\text{phys }}$, while $\eta(\cdot)$
itself is the above rigging map.
\item \textbf{Operators:} The operators act similarly to their action on
the generic $\Phi^{*}$ space; we have 
\begin{equation}
\left(\hat{O}\Psi^{\text{phys}}_{\phi}\right)[f]\coloneqq\Psi^{\text{phys}}_{\phi}\left[\hat{O}^{\prime\dagger}f\right],\quad\forall f\in\Phi,\,\Psi^{\text{phys}}_{\phi}\in\mathcal{V}_{\text{phys }}.
\end{equation}
\item \textbf{Inner product:} The physical inner product is now defined
as
\begin{align}
\left\langle \Psi^{\text{phys}}_{\phi_{1}}\vert\Psi^{\text{phys}}_{\phi_{2}}\right\rangle _{\text{phys}}= & \left\langle \eta\left(\phi_{1}\right)\mid\eta\left(\phi_{2}\right)\right\rangle _{\text{phys}}\nonumber \\
\coloneqq & \eta\left(\phi_{1}\right)\left[\phi_{2}\right]\nonumber \\
= & \frac{1}{\left(2\pi\right)^{N}}\int_{G}d\mu(g)\left\langle \hat{U}(g)\phi_{1}\mid\phi_{2}\right\rangle _{\mathrm{kin}},
\end{align}
where we demand $\eta(\phi)[\phi]\geq0$. Notice that, given that
we are working in a unitary representation of $G$ where $\hat{U}^{\dagger}\left(g\right)=\hat{U}\left(g^{-1}\right)$
and if $G$ is a unimodular group, one can also write the above as
\begin{equation}
\begin{aligned}\left\langle \Psi^{\text{phys}}_{\phi_{1}}\vert\Psi^{\text{phys}}_{\phi_{2}}\right\rangle _{\text{phys}}= & \frac{1}{\left(2\pi\right)^{N}}\int_{G}d\mu(g)\left\langle \hat{U}(g)\phi_{1}\mid\phi_{2}\right\rangle _{\mathrm{kin}}\\
= & \frac{1}{\left(2\pi\right)^{N}}\int_{G}d\mu(g)\left\langle \phi_{1}\mid\hat{U}^{\dagger}(g)\phi_{2}\right\rangle _{\mathrm{kin}}\\
= & \frac{1}{\left(2\pi\right)^{N}}\int_{G}d\mu(g)\left\langle \phi_{1}\mid\hat{U}\left(g^{-1}\right)\phi_{2}\right\rangle _{\mathrm{kin}}
\end{aligned}
\end{equation}
Now if we change the variables as $h=g^{-1}$, and since for unimodular
groups, the Haar measure is invariant under inversion, i.e., $d\mu(g)=d\mu\left(g^{-1}\right)=d\mu(h)$,
we can write this as (since $\phi_{1}$ does not depend on $h$ or
equivalently the group parameters $\lambda^{I}$)
\begin{equation}
\begin{aligned}\left\langle \Psi^{\text{phys}}_{\phi_{1}}\vert\Psi^{\text{phys}}_{\phi_{2}}\right\rangle _{\text{phys}}= & \frac{1}{\left(2\pi\right)^{N}}\int_{G}d\mu(h)\left\langle \phi_{1}\mid\hat{U}\left(h\right)\phi_{2}\right\rangle _{\mathrm{kin}}\\
= & \frac{1}{\left(2\pi\right)^{N}}\left\langle \phi_{1}\bigg|\left(\int_{G}d\mu(h)\,\hat{U}\left(h\right)\phi_{2}\right)\right\rangle _{\mathrm{kin}}\\
= & \left\langle \phi_{1}\bigg|\Psi^{\text{phys}}_{\phi_{2}}\right\rangle _{\mathrm{kin}}.
\end{aligned}
\label{eq:phys-phys-to-phys-kin}
\end{equation}
Furthermore, given that $\hat{U}=e^{-i\lambda^{I}\hat{C}_{I}},\,I=1,\ldots,N$,
we can write the second line above as
\begin{equation}
\begin{aligned}\left\langle \phi_{1}\bigg|\left(\frac{1}{\left(2\pi\right)^{N}}\int_{G}d\mu(h)\,\hat{U}\left(h\right)\phi_{2}\right)\right\rangle _{\mathrm{kin}}= & \left\langle \phi_{1}\bigg|\left(\frac{1}{\left(2\pi\right)^{N}}\int_{G}d^{N}\lambda\,\hat{U}\left(C_{I}\right)\phi_{2}\right)\right\rangle _{\mathrm{kin}}\\
= & \left\langle \phi_{1}\bigg|\left(\frac{1}{\left(2\pi\right)^{N}}\int_{G}d^{N}\lambda\,e^{-i\lambda^{I}\hat{C}_{I}}\phi_{2}\right)\right\rangle _{\mathrm{kin}}\\
= & \left\langle \phi_{1}\bigg|\left(\frac{1}{\left(2\pi\right)^{N}}\int_{G}d^{N}\lambda\,e^{-i\lambda^{I}\hat{C}_{I}}\right)\phi_{2}\right\rangle _{\mathrm{kin}}\\
= & \left\langle \phi_{1}\bigg|\delta^{N}\left(\hat{C}_{I}\right)\bigg|\phi_{2}\right\rangle _{\mathrm{kin}}
\end{aligned}
\end{equation}
\item There may be states $\phi_{\text{null }}$ that yield zero norm: $\Vert\eta\left(\phi_{\text{null }}\right)\Vert_{\text{phys}}=\eta\left(\phi_{\text{null }}\right)\left[\phi_{\text{null }}\right]=0$.
We define the null space $\mathcal{N}\subset\mathcal{V}_{\text{phys }}$
as the set of all such zero-norm states. To satisfy the axioms of
a true inner product, we quotient out this null space and form the
quotient inner product space $\mathcal{V}_{\text{phys }}/\mathcal{N}$.
The final physical Hilbert space $\mathscr{H}_{\text{phys }}$ is
defined as the Cauchy completion of this quotient space with respect
to the physical norm $\|\eta(\phi)\|_{\text{phys }}=\sqrt{\eta(\phi)[\phi]}$,
\[
\mathscr{H}_{\text{phys }}\coloneqq\overline{\left(\mathcal{V}_{\text{phys }}/\mathcal{N}\right)}.
\]
\end{enumerate}
\end{enumerate}


\section{Minimum area for real states }\label{sec:app-general-clock-min-area}

In this subsection, we will show that the result found in Sec.
\ref{subsec:Gaussian-Area} $A_{\text{min}}>0$
holds not only for the specific case when $\hat{T}=\hat{b}$, but i) for all the clocks that satisfy
$[\hat{T},\hat{H}_{C}]=i\hbar$, i.e., for a generic function $g(p_{b})$,
and ii) any real state $\psi_{a|b}(p_{a})$ \footnote{Equivalently, a state with a global and constant complex phase, i.e.
$\psi_{a|b}(p_{a})=|\psi_{a|b}(p_{a})|e^{i\alpha}$ with $\alpha=$const.} and not just the Gaussian ones.

To this end we consider the expression for $\left\langle \hat{F}_{A,T}\left(\tau\right)\right\rangle _{\text{phys}}$
from Eq.\eqref{eq:F-Ab-phys-phys-explicit} with $g\neq0$
\begin{equation}
\begin{aligned}\left\langle \hat{F}_{A,T}\left(\tau\right)\right\rangle _{\text{phys}}= & 4\pi\alpha^{4}\tau^{2}\int_{\mathbb{R}}dp_{a}\,\frac{\psi_{a|b}\left(p_{a}\right)^{2}}{p^{4}_{a}}+8\pi\alpha^{2}\tau\int_{\mathbb{R}}dp_{a}\,\frac{\partial\tilde{g}\left(p_{a}\right)}{\partial p_{a}}\frac{\psi_{a|b}\left(p_{a}\right)^{2}}{p^{2}_{a}}\\
 & +4\pi\int_{\mathbb{R}}dp_{a}\,\left\{ \hbar^{2}\left[\frac{\partial\psi_{a|b}\left(p_{a}\right)}{\partial p_{a}}\right]^{2}+\left[\frac{\partial\tilde{g}\left(p_{a}\right)}{\partial p_{a}}\right]^{2}\psi_{a|b}\left(p_{a}\right)^{2}\right\} 
\end{aligned}
\end{equation}
This is in the parabolic form in $\tau$,
\begin{equation}
\left\langle \hat{F}_{A,T}\left(\tau\right)\right\rangle _{\text{phys}}=\mathcal{A}\left(\alpha^{2}\tau\right)^{2}+\mathcal{B}\left(\alpha^{2}\tau\right)+\mathcal{C}
\end{equation}
where 
\begin{equation}
\begin{aligned}\mathcal{A}= & 4\pi\int_{\mathbb{R}}dp_{a}\,\frac{\psi_{a|b}\left(p_{a}\right)^{2}}{p^{4}_{a}},\\
\mathcal{B}= & 8\pi\int_{\mathbb{R}}dp_{a}\,\frac{\partial\tilde{g}\left(p_{a}\right)}{\partial p_{a}}\frac{\psi_{a|b}\left(p_{a}\right)^{2}}{p^{2}_{a}},\\
\mathcal{C}= & 4\pi\int_{\mathbb{R}}dp_{a}\,\left\{ \hbar^{2}\left[\frac{\partial\psi_{a|b}\left(p_{a}\right)}{\partial p_{a}}\right]^{2}+\left[\frac{\partial\tilde{g}\left(p_{a}\right)}{\partial p_{a}}\right]^{2}\psi_{a|b}\left(p_{a}\right)^{2}\right\} .
\end{aligned}
\end{equation}
Since $\mathcal{A}>0$, to show that the minimum area is strictly
positive for all $g(p_{a}) \neq 0$, a sufficient condition is to show that
the discriminant $\Delta=\mathcal{B}^{2}-4\mathcal{A}\mathcal{C}<0$,
or $\mathcal{B}^{2}<4\mathcal{A}\mathcal{C}$. However,
\begin{equation}
\mathcal{B}^{2}=64\pi^{2}\left(\int_{\mathbb{R}}dp_{a}\,\frac{\partial\tilde{g}\left(p_{a}\right)}{\partial p_{a}}\frac{\psi_{a|b}\left(p_{a}\right)^{2}}{p^{2}_{a}}\right)^{2}
\end{equation}
and using the Cauchy-Schwarz inequality $\Big(\int fg\Big)^{2}\leq\Big(\int f^{2}\Big)\Big(\int g^{2}\Big)$
for $f=\sqrt{8\pi} \frac{\psi_{a|b}\left(p_{a}\right)}{p^{2}_{a}}$ and $g=\sqrt{8 \pi} \psi_{a|b}\left(p_{a}\right)\frac{\partial\tilde{g}\left(p_{a}\right)}{\partial p_{a}}$,
we obtain 
\begin{equation}
\begin{aligned}\mathcal{B}^{2}\leq & 64\pi^{2}\int_{\mathbb{R}}dp_{a}\,\frac{\psi_{a|b}\left(p_{a}\right)^{2}}{p^{4}_{a}}\int_{\mathbb{R}}dp^{\prime}_{a}\,\left[\frac{\partial\tilde{g}\left(p^{\prime}_{a}\right)}{\partial p^{\prime}_{a}}\right]^{2}\psi_{a|b}\left(p^{\prime}_{a}\right)^{2}\\
= & 64\pi^{2}\int_{\mathbb{R}}dp_{a}\int_{\mathbb{R}}dp^{\prime}_{a}\,\frac{\psi_{a|b}\left(p_{a}\right)^{2}\psi_{a|b}\left(p^{\prime}_{a}\right)^{2}}{p^{4}_{a}}\left[\frac{\partial\tilde{g}\left(p^{\prime}_{a}\right)}{\partial p^{\prime}_{a}}\right]^{2}
\end{aligned}
\end{equation}
This means that we can also write
\begin{equation}
\begin{aligned}\mathcal{B}^{2}< & 64\pi^{2}\int_{\mathbb{R}}dp_{a}\int_{\mathbb{R}}dp^{\prime}_{a}\,\frac{\psi_{a|b}\left(p_{a}\right)^{2}\psi_{a|b}\left(p^{\prime}_{a}\right)^{2}}{p^{4}_{a}}\left[\frac{\partial\tilde{g}\left(p^{\prime}_{a}\right)}{\partial p^{\prime}_{a}}\right]^{2}+J\end{aligned}
\label{eq:B2-J}
\end{equation}
 where $J>0$ is any strictly positive expression. On the other hand we have
\begin{equation}
\begin{aligned}4\mathcal{A}\mathcal{C}= & 64\pi^{2}\int_{\mathbb{R}}dp_{a}\int_{\mathbb{R}}dp^{\prime}_{a}\,\hbar^{2}\frac{\psi_{a|b}\left(p_{a}\right)^{2}}{p^{4}_{a}}\left[\frac{\partial\psi_{a|b}\left(p^{\prime}_{a}\right)}{\partial p^{\prime}_{a}}\right]^{2}\\
 & +64\pi^{2}\int_{\mathbb{R}}dp_{a}\int_{\mathbb{R}}dp^{\prime}_{a}\,\left[\frac{\partial\tilde{g}\left(p^{\prime}_{a}\right)}{\partial p^{\prime}_{a}}\right]^{2}\frac{\psi_{a|b}\left(p_{a}\right)^{2}}{p^{4}_{a}}\psi_{a|b}\left(p^{\prime}_{a}\right)^{2}
\end{aligned}
\label{eq:4AC}
\end{equation}
Since the last line in the above is strictly positive, and by comparing
\eqref{eq:B2-J} and \eqref{eq:4AC}, we conclude that indeed $\mathcal{B}^{2}<4\mathcal{A}\mathcal{C}$
and thus $\Delta<0$. As a result, we have proven that $A_{\text{min}}>0$
is valid for any generic function $g(p_{b})$, and any real state
$\psi_{a|b}(p_{a})$.



\bibliographystyle{jhep}





\bibliography{biblio}{}

\end{document}